# Predictions for the Dynamical States of the Didymos System before and after the Planned DART Impact

Derek C. Richardson[1], Harrison F. Agrusa[1], Brent Barbee[2], William F. Bottke[3], Andrew F. Cheng[4], Siegfried Eggl[5,6,7], Fabio Ferrari[8], Masatoshi Hirabayashi[9], Özgür Karatekin[10], Jay McMahon[11], Stephen R. Schwartz[12,13], Ronald-Louis Ballouz[4], Adriano Campo Bagatin[14], Elisabetta Dotto[15], Eugene G. Fahnestock[16], Oscar Fuentes-Muñoz[11], Ioannis Gkolias[17], Douglas P. Hamilton[1], Seth A. Jacobson[18], Martin Jutzi[8], Josh Lyzhoft[2], Rahil Makadia[5], Alex J. Meyer[11], Patrick Michel[19], Ryota Nakano[20], Guillaume Noiset[10], Sabina D. Raducan[8], Nicolas Rambaux[7], Alessandro Rossi[21], Paul Sánchez[22], Daniel J. Scheeres[11], Stefania Soldini[23], Angela M. Stickle[4], Paolo Tanga[19], Kleomenis Tsiganis[17], and Yun Zhang[24,19]

[1] Department of Astronomy, University of Maryland, College Park, MD 20742, USA; dcr@umd.edu
[2] NASA Goddard Space Flight Center, Greenbelt, MD 20771, USA
[3] Southwest Research Institute, Boulder, CO 80302, USA
[4] Johns Hopkins University Applied Physics Laboratory, Laurel, MD 20723, USA
[5] Department of Aerospace Engineering, University of Illinois at Urbana-Champaign, Urbana, IL 61801, USA
[6] National Center for Supercomputing Applications, University of Illinois at Urbana-Champaign, Urbana, IL 61801, USA
[7] IMCCE, Paris Observatory 77 Avenue Denfert-Rochereau, 75014 Paris, France
[8] Space Research and Planetary Sciences, Physics Institute, University of Bern, Bern 3012, Switzerland
[9] Department of Aerospace Engineering/Geosciences, Auburn University, Auburn, AL 36849, USA
[10] Royal Observatory of Belgium, 3 Avenue Circulaire, 1180 Brussels, Belgium
[11] Smead Department of Aerospace Engineering Sciences, University of Colorado Boulder, 3775 Discovery Dr, Boulder, CO 80303, USA
[12] Planetary Science Institute, Tucson, AZ 85719, USA
[13] Lunar & Planetary Laboratory, University of Arizona, Tucson, AZ 85721, USA
[14] IUFACyT—DFISTS, Universidad de Alicante, P.O. Box 99, 03080 Alicante (Spain)
[15] INAF—Osservatorio Astronomico di Roma, 00078 Monte Porzio Catone (Roma), Italy
[16] Jet Propulsion Laboratory, California Institute of Technology, Pasadena, CA 91109, USA
[17] Department of Physics, Aristotle University of Thessaloniki, GR 54124, Thessaloniki, Greece
[18] Department of Earth and Environmental Sciences, Michigan State University, East Lansing, MI 48824, USA
[19] Université Côte d'Azur Observatoire de la Côte d'Azur, CNRS, Laboratoire Lagrange, 06304 Nice, France
[20] Department of Aerospace Engineering, Auburn University, Auburn, AL 36849, USA
[21] Istituto di Fisica Applicata "Nello Carrara" (IFAC-CNR), Sesto Fiorentino 50019, Italy
[22] Colorado Center for Astrodynamics Research, University of Colorado Boulder, 3775 Discovery Drive, Boulder, CO 80303, USA
[23] Department of Mechanical, Materials and Aerospace Engineering, University of Liverpool, Liverpool, L69 3BX, UK
[24] Department of Aerospace Engineering, University of Maryland, College Park, MD 20742, USA

## Abstract

NASA's Double Asteroid Redirection Test (DART) spacecraft is planned to impact the natural satellite of (65803) Didymos, Dimorphos, at around 23:14 UTC on 2022 September 26, causing a reduction in its orbital period that will be measurable with ground-based observations. This test of kinetic impactor technology will provide the first estimate of the momentum transfer enhancement factor $\beta$ at a realistic scale, wherein the ejecta from the impact provide an additional deflection to the target. Earth-based observations, the LICIACube spacecraft (to be detached from DART prior to impact), and ESA's follow-up Hera mission, to launch in 2024, will provide additional characterizations of the deflection test. Together, Hera and DART comprise the Asteroid Impact and Deflection Assessment cooperation between NASA and ESA. Here, the predicted dynamical states of the binary system upon arrival and after impact are presented. The assumed dynamically relaxed state of the system will be excited by the impact, leading to an increase in eccentricity and a slight tilt of the orbit, together with enhanced libration of Dimorphos, with the amplitude dependent on the currently poorly known target shape. Free rotation around the moon's long axis may also be triggered, and the orbital period will experience variations from seconds to minutes over timescales of days to months. Shape change of either body, due to cratering or mass wasting triggered by crater formation and ejecta, may affect $\beta$, but can be constrained through additional measurements. Both BYORP and gravity tides may cause measurable orbital changes on the timescale of Hera's rendezvous.

*Unified Astronomy Thesaurus concepts:* Asteroid dynamics (2210); Asteroids (72); Asteroid satellites (2207)

## 1. Introduction

The Double Asteroid Redirection Test (DART) is a NASA mission that will demonstrate the use of a kinetic impactor for defense against objects on a collision course with Earth (Cheng et al. 2018; Rivkin et al. 2021). Following its successful launch on 2021 November 24, at 06:21:02 UTC, the DART spacecraft is planned to impact (65803) Didymos I Dimorphos, the satellite of the (65803) Didymos binary system, on 2022 September 26, at approximately 23:14 UTC, causing a minimum 73 s change in the binary orbital period that will be measurable with Earth-based observations. The impact-induced change in period will be determined by the momentum of the spacecraft and the fate of any resulting ejecta; the momentum transfer enhancement factor "Beta" ($\beta \geq 1$) will characterize any additional kick given to Dimorphos as a result of escaping ejecta and will be estimated from observations during the event (Rivkin et al. 2021). The Italian Space Agency's Light Italian CubeSat for Imaging of Asteroids (LICIACube) aboard DART will be deployed 10 days before impact, complementing inbound imagery from the spacecraft by recording the impact event and ejecta evolution and by obtaining the shape of Dimorphos opposite the impact site (Dotto et al. 2021). The European Space Agency mission Hera consists of an orbiter and two CubeSats that will visit the Didymos system 4 yr after the DART impact, to fully characterize the composition, surface, and interior structures, as well as the dynamical states of the bodies, and to assess further the impact effects (Michel et al. 2022). Together, DART and Hera are supported by the Asteroid Impact and Deflection Assessment (AIDA) cooperation between the two space agencies.

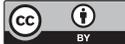

In this paper, predictions for the dynamical states of Didymos before and after the DART impact are presented, as these will be needed for the interpretation of the $\beta$ measurement and as a check on predictions of the impact outcome and expected observations. Section 2 summarizes current knowledge of the Didymos system's dynamical state, along with the implications of any uncertainties, including whether the system is likely in a dynamically relaxed state. The numerical code adopted for characterizing the rigid-body dynamics both pre- and post-impact is also presented. In Section 3, the expected dynamical effect of the impact as a function of $\beta$ is discussed in the rigid-body limit. This includes a discussion of the dynamical instabilities that could be triggered in Dimorphos, depending on its actual shape. Section 4 summarizes the preliminary findings from models that relax the rigid-body assumption by replacing Dimorphos with a rubble pile. In Section 5, the implications for the $\beta$ measurement of any body shape change induced by the impact on the secondary or accretion of ejecta on the primary are presented. Section 6 rounds out the discussion by summarizing the secular effects that may be measurable over the time frames of DART and Hera, specifically orbital evolution driven by BYORP and/or gravitational tides. The conclusions of this study and the future outlook are given in Section 7, along with a summary table of the dynamical effects considered here and a figure illustrating them. Finally, the reader is referred to the companion papers in this special issue to form a complete picture of the expectations for the DART mission.

## 2. The Didymos System's Current Dynamical State

The current dynamical state of the Didymos system is informed by both the observations conducted to date and the numerical simulations based on those observations. The simplest characterization of the system is to treat it as a classic gravitational two-body problem. This approach, augmented by a secular term to account for radiation-induced acceleration (BYORP—see Section 6), is sufficient to predict the location of Dimorphos on its orbit around Didymos with the required precision for spacecraft targeting (Rivkin et al. 2021). However, Didymos has an irregular top-like shape, Dimorphos is assumed to have a slightly elongated shape, and both bodies are in close proximity to one another, making non-Keplerian terms in the equations of motion significant, both with respect to predicting any possible excited dynamical states prior to the DART impact and for the development of any excited modes post-impact. Since this complication may have a bearing on the pre-impact configuration and the post-impact measurements, it is necessary to model these effects to assess their importance. The most straightforward approach is to assume rigid shapes for the components, based on the best-available measurements, and to model the coupled rigid-body motion using a numerical integrator. This approach is described in this section. In later sections, the rigid-body assumption is relaxed, since Didymos or Dimorphos or both may be rubble piles (e.g., Walsh & Jacobson 2015), but since many-body simulations are far more expensive to compute, those studies are presently limited in scope.

### 2.1. Origin

Didymos is a near-Earth asteroid (NEA) with semimajor axis, eccentricity, and inclination ($a$, $e$, $i$) values of (1.64 au, 0.384, 3.4°). It has an orbit that just crosses the Earth's path at perihelion. Such values are common for bodies that have recently escaped the main asteroid belt via a resonance (e.g., Bottke et al. 2002). Using the NEA population models from Bottke et al. (2002), Granvik et al. (2016), and Granvik & Brown (2018), combined with the dynamical calculations used in Bottke et al. (2015), it is possible to make predictions of where Didymos may have come from in the main belt.

The NEA model from Bottke et al. (2002) assumes that objects with $a < 7.4$ au and absolute magnitude $H < 22$ primarily come from the main asteroid belt or the trans-Neptunian region. More specifically, in this model, NEAs are derived from one of five primary source regions: the $\nu_6$ secular resonance along the inner edge of the main belt, the intermediate-source Mars-crossing region adjacent to the main belt (IMC), the 3:1 mean-motion resonance with Jupiter at 2.5 au, the outer main belt region beyond 2.8 au, and the Jupiter-family comet region, a zone that is resupplied by objects from the scattered disk in the trans-Neptunian region. The NEA model from Granvik et al. (2016) and Granvik & Brown (2018) uses a methodology similar to Bottke et al. (2002), but is more advanced in several ways. They divide the NEA source regions of the main asteroid belt into six different entrance routes to the planet-crossing region: the $\nu_6$ secular resonance, the 3:1, 5:2, and 2:1 mean-motion resonance complexes with Jupiter, and the Hungaria and Phocaea small-body regions adjacent to the main belt.

In both models, the dynamical pathways taken by test bodies from the source regions are tracked across a network of cells in ($a$, $e$, $i$) space. The length of time spent by a particle in each cell is tabulated, yielding a residence time probability distribution for each source. These sources are then combined with weighting functions and observational selection effects in order to fit the net function to the observed NEAs. This yields an estimate of the present-day steady-state orbital distribution of the NEA population. The probability that an NEA came from a given source region can then be estimated using its ($a$, $e$, $i$) values, with that information provided by the model results within each cell.

Starting with the more advanced NEA model from Granvik et al. (2016) and Granvik & Brown (2018), Didymos likely reached its current orbit by exiting the inner main belt near or within the $\nu_6$ resonance between 2.1–2.5 au (>82% chance). Other possible source regions are the Hungaria asteroids (8%) and the inner/central main belt via the 3:1 mean-motion resonance with Jupiter (7%). The Bottke et al. (2002) model yields similar results: it suggests that Didymos has a 55% chance of coming from the $\nu_6$ resonance, a 36% chance of coming from the IMC region, and a 9% chance of coming from the 3:1 resonance.

Another way to gain more specificity about the origin location of Didymos is to use the method of Bottke et al. (2015). They tracked the orbital evolutions of test asteroids that evolved into the $\nu_6$ resonance, the IMC region, and the 3:1 resonance by Yarkovsky thermal drift. For those test asteroids that reached the planet-crossing region, they determined which ones passed very close to the current $(a, e, i)$ orbit of Didymos. Here, a good match was arbitrarily defined as one with $\Delta a \leqslant 0.01$ au, $\Delta e \leqslant 0.01$, and $\Delta i \leqslant 1°$. Some test bodies also met this threshold at more than a single time step interval. Weighting these test body results by the source probability results from above, Didymos most likely exited the main belt with semimajor axis $a \leqslant 2.2$ au and inclination $1° < i < 6°$. For the latter, the most favored values for the test bodies are between $4° < i < 5°$ (nearly 30% of all bodies).

Remote observations show that Didymos is an S-type asteroid that is spectroscopically most consistent with ordinary chondrites, with an affinity for L/LL-type meteorites (Dunn et al. 2013). Didymos also originated from a relatively high-albedo parent asteroid or parent family, with its geometric albedo being $0.15 \pm 0.04$ (Naidu et al. 2020). This value is consistent with the mean albedo of the prominent Baptistina family in the inner main-belt region, 0.16 (Nesvorný et al. 2015). However, additional family candidates may also be plausible parent families, such as Flora (mean albedo 0.30), Nysa/Hertha (mean albedo 0.28), Massalia (mean albedo 0.22), and Lucienne (mean albedo 0.22) (Nesvorný et al. 2015). It is also possible that Didymos was derived from a smaller family or an immediate precursor that was not associated with a family. More work will be needed to explore these possibilities.

Using the predicted semimajor axis and inclination values that Didymos had when it left the main belt, a few candidate families can potentially be ruled out. For example, Lucienne family members have a mean inclination of 12° (Nesvorný et al. 2015). Objects in a family will drift inward and outward by Yarkovsky thermal drift, with some reaching resonances that can transport them out of the asteroid belt. The inclinations they have when they enter those resonances, however, should be similar to those they had in the family. This implies that Lucienne family members are unlikely to reach the current orbit of Didymos (with $1° < i < 6°$), and therefore this family can be ruled out as a plausible source of the DART target. As a second example, the Massalia family formed less than 200 Myr ago (Vokrouhlický et al. 2006), and Didymos-sized family members have only drifted far enough by the Yarkovsky effect in that time to escape the main belt via the 3:1 resonance. This source region is not favored by any of the models and is inconsistent with the favored departure semimajor axis values of $a \leqslant 2.2$ au.

Escaping members from the Baptistina, Flora, and Nysa/Hertha families can plausibly reproduce the predicted semimajor axis and inclination values of Didymos. All must be considered potential source families. However, the mean inclination of the Nysa/Hertha members is 2.5°, while those of Baptistina and Flora are 5.6° and 5.1°, respectively. Given that the most favored inclination values for Didymos when departing the main belt are $4° < i < 5°$, Baptistina/Flora family members would more easily hit this dynamical "sweet spot" than Nysa/Hertha.

Finally, when the mean albedos of Baptistina/Flora are included with the above constraints, it seems that Baptistina is the most likely source of all of the inner main-belt families considered here. The DART and Hera investigations of Didymos will hopefully provide additional diagnostic data that can be used to further test the origin of Didymos.

**Table 1**
Selected Dynamical Parameters from Rivkin et al. (2021)[a]

| Parameter | Value |
|---|---|
| Volume-equivalent Diameter of Primary $D_P$ | $780 \pm 30$ m |
| Volume-equivalent Diameter of Secondary $D_S$ | $164 \pm 18$ m |
| Bulk Densities of Components $\rho_P$ | $2170 \pm 350$ kg m$^{-3}$ |
| Mean Separation of Component Centers $a_{\rm orb}$ | $1.20 \pm 0.03$ km |
| Secondary Shape Elongation $a_S/b_S$, $b_S/c_S$ | $1.3 \pm 0.2$, $1.2$ |
| Total Mass of System $M$ | $(5.55 \pm 0.42) \times 10^{11}$ kg |
| Secondary Orbital Period $P_{\rm orb}$ | $11.921\,628\,9 \pm 0.0000028$ hr |
| Secondary Orbital Eccentricity $e_{\rm orb}$ | $<0.03$ |
| Primary Rotation Period $P_P$ | $2.260\,0 \pm 0.0001$ hr |
| Secondary Rotation Period $P_S$ | $P_{\rm orb}$ (assumed to be tidally locked) |
| Secondary Orbital Inclination $i_{\rm orb}$ | $0°$ (assumed) |

**Note.**
[a] Updated with the latest values as of this writing (DRA v. 3.2).

### 2.2. Measured and Inferred Dynamical Parameters

Table 1 summarizes the important dynamical parameters of the Didymos system from (Rivkin et al. 2021, Appendix A), updated with the latest values as of this writing from the Design Reference Asteroid v. 3.2 (DRA; DART mission internal document). These parameters are either measured or inferred from ground-based observations. The most uncertain parameters are the eccentricity and inclination of the binary orbit, the shape of Dimorphos, and the amplitudes of any excited dynamical modes, such as libration (assumed to be zero in the table). The shape uncertainty in particular leads to a wide parameter space of possible dynamical states both before and after impact, as discussed below. The values from the table and other sources, including the adopted shape models (Naidu et al. 2020), are used to set the initial conditions for the numerical simulations of the system, discussed next.

### 2.3. Full Rigid Two-body Modeling

Due to the close proximity and nonspherical shapes of Didymos and Dimorphos, their translational and rotational states are highly coupled, leading to non-Keplerian motion. Full Two-body Problem (F2BP) codes can be used to capture the evolution of the mutual orbit and body spin states (Fahnestock & Scheeres 2006) under the assumption that the bodies are rigid. The F2BP approach was applied to the Didymos system by Agrusa et al. (2020) to identify the degree of spin–orbit coupling between the bodies, to understand the system's sensitivity to initial conditions, and to select the best-suited F2BP code for the DART investigation. Due to the asymmetric shape of Didymos, they found that the simulated mutual orbit period was sufficiently sensitive to Didymos's initial rotation phase that predictions of Dimorphos's true anomaly at later times were impractical to obtain with numerical methods. They also found that the libration amplitude of Dimorphos is highly dependent on changes to its velocity; that is, small perturbations (i.e., the DART impact) should strongly affect the resulting libration state of Dimorphos. Finally, they determined that the General Use Binary Asteroid Simulator (GUBAS) was the best choice in terms of both accuracy and speed for studying the Didymos binary, making it the adopted simulation code for DART-related rigid-body studies of the Didymos system. The simulations described in this section below use GUBAS exclusively. The code is available as a free download[25]—it is a simple, fast, and open-source tool for modeling the coupled rotational and translational dynamics of two bodies with arbitrary mass distributions (Davis & Scheeres 2020). The code has since been modified to include third-body perturbations (Meyer & Scheeres 2021) and tidal dissipation (Ö. Karatekin 2022, in preparation).

### 2.4. Relaxed-state Assumption

Ground-based observations indicate that the eccentricity of Didymos's mutual orbit is consistent with zero, i.e., a circular orbit, although an upper limit as high as 0.05 has been reported (Scheirich & Pravec 2009; Fang & Margot 2012; Naidu et al. 2020); the adopted upper limit is 0.03 (Table 1). Furthermore, Didymos's fast rotation and top-like shape are indicative of a possible rubble-pile structure (Richardson & Walsh 2006)—this motivates the rubble-pile studies discussed later in this paper, but treating the bodies as rigid for the purpose of integrating the dynamics from a relaxed state is a good first approximation. Due to the highly dissipative nature of rubble-pile asteroids, it is likely that mutual tides will have damped the system to an equilibrium state with Dimorphos, in synchronous rotation and a near-circular mutual orbit (Goldreich & Sari 2009). For these reasons, it is assumed that the pre-impact state of the system is relaxed, and that the DART impact will instantaneously alter Dimorphos's orbital velocity, leading to excitement of both the mutual orbit and Dimorphos's spin state. If, upon DART's arrival, it is found that the system is already dynamically excited, then this assumption will need to be reevaluated. Effectively, the "relaxed-state assumption" provides a lower bound on the expected changes to the Didymos system resulting from DART.

Due to the non-Keplerian nature of the Didymos system's binary dynamics, initializing a dynamically relaxed state for simulations is nontrivial. The best-constrained parameters for the system from observations are the mutual orbit period and the component separation or orbital semimajor axis (Table 1). Therefore, the priority is generating initial conditions that yield a circular orbit with Dimorphos in synchronous rotation that also match the measured orbit period and semimajor axis. The approach is to assume that both Didymos and Dimorphos have the same density—which is plausible if they have the same origin—and then the total mass $M$ is adjusted in GUBAS simulations until the mutual orbit period matches within some specified level of precision ($M$ in Table 1 assumes a Keplerian orbit). The adopted algorithm uses a secant search method and converges within a few short GUBAS runs. Because the mutual orbit period is so sensitive to the mass distribution, this procedure must be redone whenever the shape of Didymos or Dimorphos is refined, or whenever any other initial condition parameters are changed. The details of the algorithm are discussed in Agrusa et al. (2021).

It is possible that Didymos and Dimorphos have different bulk densities, which this relaxation algorithm overlooks. If Dimorphos formed through YORP spin-up, followed by mass loss and subsequent gravitational accumulation in orbit, then it likely has a similar bulk density to that of Didymos. However, there is also the possibility that Dimorphos formed as a monolith from a single fission event, in which case its bulk density may be different from that of Didymos (Walsh & Jacobson 2015). For now, to keep the simulation parameter space manageable, we assume that both bodies have the same bulk density. Following DART's arrival, it will be possible to place better constraints on Dimorphos's bulk density via some combination of direct images taken with DART and through impact simulations (Daly et al. 2022; Stickle et al. 2022). However, strong constraints on Dimorphos's bulk density, as well as its origin, will not be available until the arrival of Hera (Michel et al. 2022).

### 2.5. Possible Sources of Pre–DART impact Excitation

Although the relaxed-state assumption is well motivated (Pravec et al. 2016 found that out of 29 close binary systems, only five were asynchronous), there are several possible sources of the dynamical excitation of the system that could result in a perturbation that is not fully damped by the time of DART's arrival. For example, Meyer & Scheeres (2021) demonstrated that a relaxed binary system can be pushed into a chaotic state by a close planetary flyby, which could lead to the tumbling of one or both of the components. Quillen et al. (2022) predict that out-of-plane rotation in the secondary can be long lasting, and Ćuk et al. (2021) also show that out-of-plane rotation can have significant effects on the dynamics. Other methods of excitation include solar tides, such as in the case of 66391 Moshup (provisional designation 1999 KW$_4$; Scheeres et al. 2006), and previous natural impacts on the secondary (Yanagisawa 2002).

As a brief analysis of the likelihood of finding Didymos in an excited state resulting from a planetary encounter, the findings in Meyer & Scheeres (2021) can be applied to Didymos. From the analytic formulae, a flyby that induces a 30° libration will also excite the eccentricity by about 0.1, on average. Considering the mean variation in eccentricity during a flyby, the heliocentric orbit of Didymos can be propagated, while keeping track of how often these flybys occur. Figure 1 shows that the probability of recently having experienced such a flyby is low. Per the propagation of 1000 virtual Didymos systems using the model of Fuentes-Muñoz et al. (2022; also see Fuentes-Muñoz & Scheeres 2020), there is less than a 5% chance of such a flyby having occurred in the last 100,000 yr.

Using the dissipation model in Goldreich & Sari (2009) and the tidal quality model for rubble piles in Nimmo & Matsuyama (2019; giving $Q/k \approx 10^6$ for the primary and $Q/k \approx 25,000$ for the secondary), an eccentricity of 0.1 is estimated to dissipate within 20,000 yr. This relatively fast dissipation time is due to the close proximity between Didymos and Dimorphos, as well as the highly dissipative nature of rubble piles. Note that there is considerable uncertainty around the tidal quality factor, so this estimate is only a first approximation. Also, if Dimorphos is a rigid body, the dissipation times will be much longer. Nevertheless, the combination of infrequent energetic flybys with fast dissipation

---

[25] https://github.com/alex-b-davis/gubas

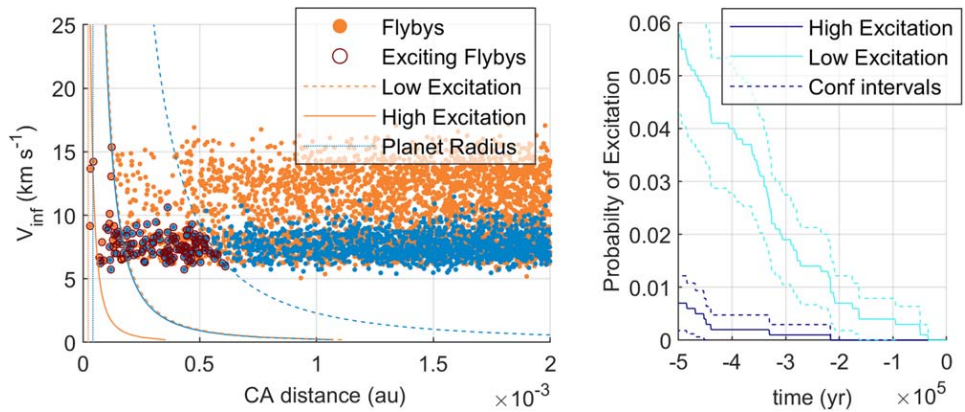

**Figure 1.** Left: potentially exciting flybys of 1000 virtual Didymos clones over the last 500,000 yr. Planetary flybys are recorded with Mars (orange) and Earth (blue), with their respective thresholds for excitation in the space of the closest approach distance and relative speed of the flyby. Right: estimated probability of experiencing a close flyby that excites Didymos, as characterized by a variation in eccentricity of $\Delta e = 0.1$ for low excitation and $\Delta e = 1$ for high excitation.

times indicates that Didymos is unlikely to be in a perturbed state from planetary encounters.

Using the same modified GUBAS code as Meyer & Scheeres (2021), a perihelion passage of Didymos can be simulated simply by substituting the Sun in place of a planet. Given the relatively distant perihelion distance ($\approx 1.01$ au), perihelion passage provides negligible excitation to the mutual orbit, eliminating this as a possible source of pre-impact excitation.

Another possible source of system excitation is meteorite impacts. As Didymos has an eccentric orbit that crosses the inner main belt (beyond 2.1 au from the Sun) for approximately 250 days out of its 770 day orbital period, a mixed-impactor population model was adopted, which includes both near-Earth objects (NEOs) and main-belt asteroids (MBAs), to estimate the impact rate on the binary system from natural impacts. While the number of potential impactors in the main asteroid belt is higher than in near-Earth space, the encounter speeds in the main belt are generally lower by a factor of approximately 4. For NEO impacts, the impactor flux model of Brown et al. (2002) is used, scaling for target size and neglecting gravitational focusing. For MBA impacts, the impact flux model of Bottke et al. (2005) is used; note, however, that the MBA population of impactors in the size range of interest (a few centimeters to decameters) is poorly constrained. Indeed, considering main-belt population models that show a depletion in small m-scale impactors (Cibulková et al. 2014; Zain et al. 2020), the NEO impact flux would dominate over the MBA flux. This is further complicated by the fact that NEA orbits are chaotic. Nevertheless, it is found that the time interval between the impacts of a given energy can be constrained by considering the NEO impact flux model of Brown et al. (2002) and the MBA impact flux model of Bottke et al. (2005).

A simplifying assumption is made in the model that the encounter speeds in the NEO region and the inner main belt occur at 18.5 km s$^{-1}$ and 5.2 km s$^{-1}$, respectively. The encounter speeds by NEAs in the inner main belt may in fact be larger (Michel et al. 1998; Dell'Oro et al. 2011). Therefore, the calculated time intervals for impacts on the Didymos system in near-Earth space are considered to be conservative. The time intervals for impacts of a given energy are shown in Figure 2 for Didymos and Dimorphos using this mixed-impactor population model (the solid and dashed black curves) and for an NEO impact–only model (the solid and dashed blue curves). Considering the aforementioned uncertainty in the population of meter-scale MBAs, the black and blue curves for each asteroid bracket the possible time intervals between the impacts of a given energy. The largest-impact energy event that Dimorphos experiences at least as frequently as the $\sim$20,000 yr damping timescale is $5 \times 10^5$ to $8 \times 10^6$ J. Given that this is less than 0.01% of the DART impact energy, which is predicted to increase the eccentricity of the system $\ll 0.1$ (Section 3.1), the effects of such a meteorite impact would damp out long before the next comparable impact. Thus, accounting for both planetary encounters and impacts, it is likely that the system is currently in a low-excitation state.

The DART impact energy is expected to be about $10^{10}$ J (Rivkin et al. 2021). An impact on Didymos of this energy is expected once every 0.33 to 7 Myr, so Didymos may have been hit many times at the DART impact energy within its median dynamical lifetime of $10^7$ yr (Gladman et al. 2000). For Dimorphos, the impact interval is much longer, due to its smaller size (8 to 159 Myr), experiencing up to one impact with an energy equivalent to DART over its lifetime. Dimorphos's specific catastrophic disruption energy ($Q_D^\star$) is roughly 200 J kg$^{-1}$ if the asteroid is a monolithic object with a significant tensile strength ($\gtrsim$1 MPa; Jutzi et al. 2010). On the other hand, if Dimorphos has a low tensile strength, $Q_D^\star$ is roughly 20 J kg$^{-1}$ (Jutzi 2015; Raducan & Jutzi 2022). This implies an impact energy at catastrophic disruption in the range of $\sim 10^{11}$–$10^{12}$ J. The DART impact energy is therefore only a factor of 10–100 smaller than Dimorphos's specific catastrophic disruption energy, so the DART impact can represent a major shattering event. At the size scale of Didymos, the effect of tensile strength on $Q_D^\star$ is negligible (Jutzi 2015). For Didymos, $Q_D^\star \simeq 500$ J kg$^{-1}$ for a catastrophic disruption energy of nearly $3 \times 10^{14}$ J, many orders of magnitude above the DART impact energy.

The DART impact on Dimorphos could cause seismic accelerations that greatly exceed the local surface gravity globally (Cheng et al. 2002). If the impact creates a crater greater than $\sim 0.1$ times the size of Dimorphos, it may cause erasure of all pre-existing craters of its size or smaller (Asphaug 2008). In order to obtain an estimate of the seismic accelerations expected by the impact, the model of Richardson et al. (2005) was used, with a seismic efficiency factor, $\eta$, of at most $10^{-6}$, as measured for the Small Carry-on Impactor (SCI) impact on asteroid Ryugu (Nishiyama et al. 2021). This leads to the following estimate of the surface acceleration, $a$, generated by global seismicity, normalized by Dimorphos's gravity, $g$, and neglecting the attenuation of the seismic wave:

$$\frac{a}{g} = 10^5 \sqrt{\frac{108}{\pi} \frac{\eta f^2}{G^2 \rho_P^3 D_S^5}}, \quad (1)$$

where $f$ is the characteristic seismic frequency of the bodies, which is taken to be on the order of 1 Hz. Equation (1) suggests global shaking of 300 × g for the DART impact on Dimorphos, and even 100 × g for a similar impact on Didymos. However, it is unclear if such levels of global acceleration will indeed be achieved, since the surface waves have been shown to attenuate rapidly on rubble piles. Honda et al. (2021) showed that the seismic wave that propagated from the formation of the SCI crater decayed at 4 crater radii.

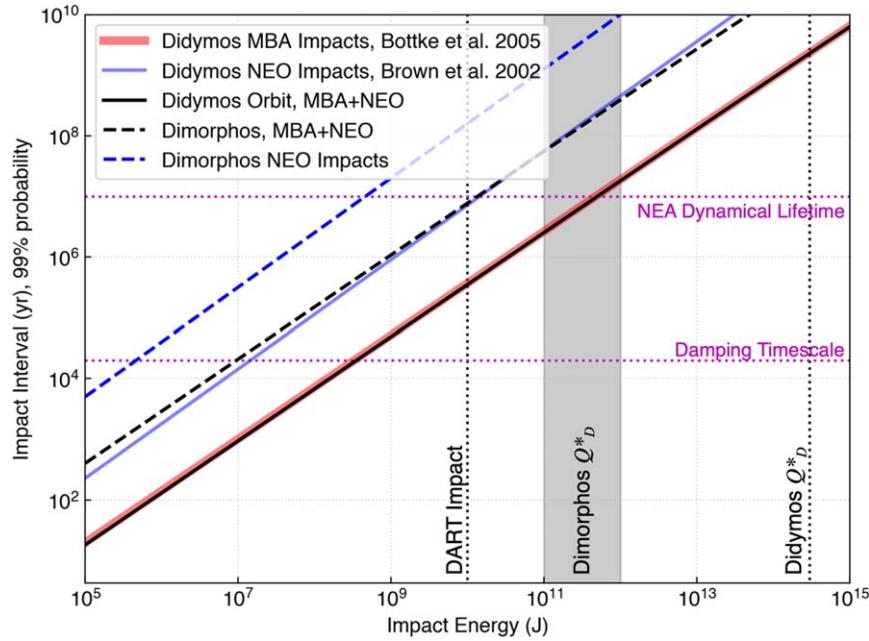

**Figure 2.** Time intervals between the impacts of a given energy for Didymos and Dimorphos. The blue and red lines show the impact intervals if Didymos were impacted by only the NEO and MBA populations, respectively. The black solid line shows the impact interval on Didymos for a mixed-population model that takes into account the system's current orbit, which crosses the inner main belt. The black dashed line shows the equivalent impact timescales for Dimorphos in the mixed-population model. The dotted magenta lines give the estimated damping timescale for a binary eccentricity of 0.1 (bottom line) and the median dynamical NEA lifetime (top line). The vertical dotted lines indicate the different impact energies of interest. The range of specific catastrophic disruption energies for Dimorphos is represented by the gray area.

### 2.6. Simulation Pipeline

The version of GUBAS used for mission data analysis is based on the public version (Section 2.3), but has been modified in several ways and is maintained at NASA's Goddard Space Flight Center (GSFC). It has been augmented to handle nonconvex shape models, updated to use PYTHON 3, paired with a BASH shell script system, enabling parallel processing of multiple simulation scenarios, and modified to output only binary data files, in order to the improve overall processing speed. Furthermore, the dynamical relaxation algorithm (Section 2.4) has been added to the scripts, so that the initial conditions can be modified to be dynamically relaxed prior to simulation execution, when necessary. This version has been augmented to allow for the primary and secondary bodies to have different bulk densities.

The Didymos system DRA data products (Section 2.2) include shape model files for both Didymos and Dimorphos, as well as ephemeris data files describing the mutual translational motion of Didymos and Dimorphos about their barycenter. The shape model for Didymos is derived from resolved radar observations (Naidu et al. 2020). However, such radar-derived shape information for Dimorphos is not available and so an ellipsoidal shape model is used, which conforms to the observational constraints on the Dimorphos/Didymos diameter ratio and observations of ellipsoidal semi-axis ratios for other binary asteroid system members. The Dimorphos shape model volume also yields a mass that conforms to the barycentric positions at the impact epoch from the ephemeris data files and the associated system mass parameters and body bulk densities shown in Table 1.

The heliocentric motion ephemeris for the Didymos system barycenter and the relative motion ephemeris for Didymos and Dimorphos relative to the system barycenter are provided by NASA/Jet Propulsion Laboratory.[26] The ephemeris for the Didymos system barycenter is produced using high-fidelity dynamics modeling that includes point-mass gravity and relativistic perturbations from appropriate solar system bodies. The Yarkovsky effect is also included, in which thermal energy from the Sun is absorbed by the rotating asteroid and then emitted anisotropically, producing net forces on the asteroid. For the motion of the bodies relative to their barycenter, Dimorphos is treated as a point mass and modified Keplerian dynamics are used that include a model for mean-motion drift due to BYORP. The model is appropriately constrained to agree with observational data.

The DRA is generally updated when observations (prior to or during the mission) yield new or updated information about the Didymos system. In response to such DRA updates for parameters that affect the dynamics simulations, a pipeline procedure for updating the dynamics simulations is executed, described next. The following items are obtained from the DRA data set maintained on the DART Science Operations Center (SOC): Dimorphos and Didymos shape model files; system mass parameters; and Didymos system ephemeris data files. The impact epoch associated with the date on which DART launched is utilized as the initial epoch for the simulation. The Didymos and Dimorphos barycentric-state vectors at that epoch are retrieved from the ephemeris data files and combined with the shape model files (after some necessary file format conversions and principal axis frame transformations) in the dynamical-state relaxation algorithm. This produces dynamically relaxed initial conditions for the system at the chosen initial epoch. Those initial conditions are then documented in the SOC. Simulations of the unperturbed system may then be performed. For simulations of the DART impact effects, changes are superimposed onto the aforementioned relaxed initial conditions, using the DART spacecraft's mass at impact, the impact's relative velocity vector, and the impact's location on the surface of Dimorphos.

---

[26] Available at https://ssd.jpl.nasa.gov/ftp/eph/small_bodies/dart/dimorphos/.

The dynamics simulations performed at GSFC are currently executed on a 192-thread system (96 Intel Xeon Gold 6252 CPUs at 2.1 GHz, two threads per core). On that system, simulating the translational and rotational motion of the Didymos system for one day at a 40 s step size requires approximately 30 to 60 s of run time. The results from a nominal simulation execution are presented in Section 3.1.

## 3. Prediction of Post-impact Dynamical State (Rigid-body Model)

The fundamental idea behind using a kinetic impactor to mitigate a small-body hazard is to modify the target orbit by imparting momentum. A key uncertainty is that an unknown amount of ejecta may be liberated from the target by the impact, enhancing the momentum transfer by a factor of magnitude $\beta > 1$. Rivkin et al. (2021) provide a formula (their Equation (1)) for calculating this momentum enhancement based on DART mission observables, given certain assumptions. Uncertainty in Didymos' mass dominates the accuracy of this calculation. Based on impact simulations, Stickle et al. (2022) find $\beta$ plausibly ranges from 1 to 5 shortly after impact, depending on details of the target properties and impact circumstances. The long-term fate of the ejecta also plays a role (Fahnestock et al. 2022) and determines the "heliocentric $\beta$," i.e., the final deviation of the system orbit relative to the Sun (Section 3.4). However, to investigate the post-impact dynamics, it is sufficient to consider a range of $\beta$ values for the orbit of Dimorphos relative to Didymos, under the assumption the momentum is transferred instantaneously. This is the approach taken below, although the effect of a non-instantaneous transfer is discussed briefly in Section 3.5.

### 3.1. The Nominal Case

The NASA/GSFC GUBAS simulation software was used to model the post-impact dynamical evolution of the Didymos system using dynamically relaxed initial conditions representing the nominal pre-impact state. Three values of $\beta$ were used: 1.0, 1.2, and 2.0. The simulation results are summarized in Table 2. Dimorphos is modeled as an ellipsoid with semi-axis lengths of $103.79 \times 79.84 \times 66.53$ m, yielding semi-axis ratios of $a/b = 1.3$ and $b/c = 1.2$. This Dimorphos shape model is

**Table 2**
Post-impact Dynamics Simulation Results for 2 Months after Impact

| $\beta$ | $\Delta v$ (mm s$^{-1}$) | $\overline{\Delta P}$ (s) | $\bar{e}$ | $\overline{\Delta i}$ |
|---|---|---|---|---|
| 1.0 | 0.734 3 | −529.2 | 0.007 4 | 0°.022 |
| 1.2 | 0.880 7 | −630.0 | 0.008 9 | 0°.029 |
| 2.0 | 1.466 9 | −1033.2 | 0.014 8 | 0°.053 |

**Note.** $\overline{\Delta P}$ represents the mean orbit period change, $\bar{e}$ represents the mean post-impact orbit eccentricity, and $\overline{\Delta i}$ represents the mean orbit plane change.

the nominal DRA-compliant shape, per the constraints and processes discussed in Section 2.6. The dynamical relaxation algorithm yields bulk densities for Didymos and Dimorphos of approximately 2201 kg m$^{-3}$ and 2202 kg m$^{-3}$, respectively, which are well within the 1$\sigma$ uncertainty of the nominal bulk density shown in Table 1.

As shown in Table 2, DART impacts in a retrograde sense, which serves to reduce the binary system orbit period by approximately 500 to 1000 s, depending on $\beta$. That amount of period change is much larger than the minimum period change requirement of 73 s. The changes in mutual orbit eccentricity and plane orientation are quite modest and are not expected to interfere with the detection of the orbit period change via remote observations. The libration angle (not shown in Table 2; see Section 3.2 for further discussion) seen in the simulation results for the pre-impact dynamics is approximately ±0.4°–0.5°. The libration angle extent increases in the post-impact simulation results to ±6.9°–13.6° for the range of $\beta$ values used (1.0–2.0). The results in Table 2 show that the imparted $\Delta v$ and the resulting changes in orbit period and eccentricity seem to scale approximately linearly with $\beta$, all else being equal. However, the change in the orbit plane orientation appears to vary more than linearly with $\beta$. The pre-impact and post-impact behaviors of the mutual orbit period and secondary body libration angle across time are shown in Figure 3, while the behavior of the orbit eccentricity is shown in Figure 4.

Simulations of off-nominal impacts were also performed using a previous version of the DRA in combination with a set of nine irregular shape models for Dimorphos that are not DRA compliant (plus one spherical shape model), off-center impact locations, ranging from only somewhat off the center to near the edge of the body, and a range of $\beta$ values. The irregular shape models were made by scaling various known asteroid shape models to the approximate size scale of Dimorphos. The masses of some of those Dimorphos models are lower than the nominal Dimorphos mass, because some of the shape models used have smaller volumes than the volumes yielded by the Dimorphos shape constraints in the DRA. In our simulations, these lower-mass Dimorphos models experience greater $\Delta v$ and more significant dynamical perturbations from the DART impact than expected for the nominal Dimorphos mass.

The imparted $\Delta v$ ranged from 0.523 mm s$^{-1}$ to 12.7 mm s$^{-1}$, the mean period change ranged from −328 s to −6460 s, the post-impact mean eccentricity ranged from 0.008 52 to 0.096 6, and the change in mean orbit plane ranged from 0.003 27° to 0.428°. These results show that even a very off-center impact into a rather irregularly shaped Dimorphos should cause a significant period change well in excess of the minimum required amount. However, the larger values of post-impact eccentricity could adversely affect the accuracy of remote

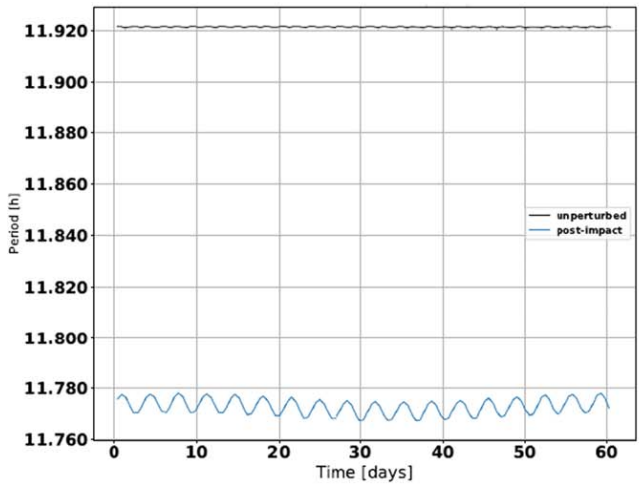
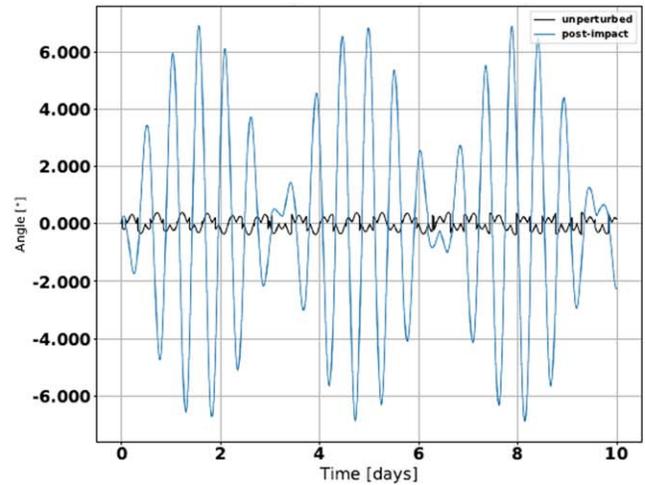

(a) $\beta = 1$, pre-impact (unperturbed) and post-impact orbit period.

(b) $\beta = 1$, pre-impact (unperturbed) and post-impact libration angle. The libration amplitude pre-impact is $< 0.4°$.

**Figure 3.** GUBAS simulation results showing the orbit period and libration angle behavior, pre-impact and post-impact. The behaviors of these quantities remain consistent over longer time spans, e.g., out to 1 yr. Only 10 days of propagation are shown here for the libration angle, so that the behavior is more readily discernible.

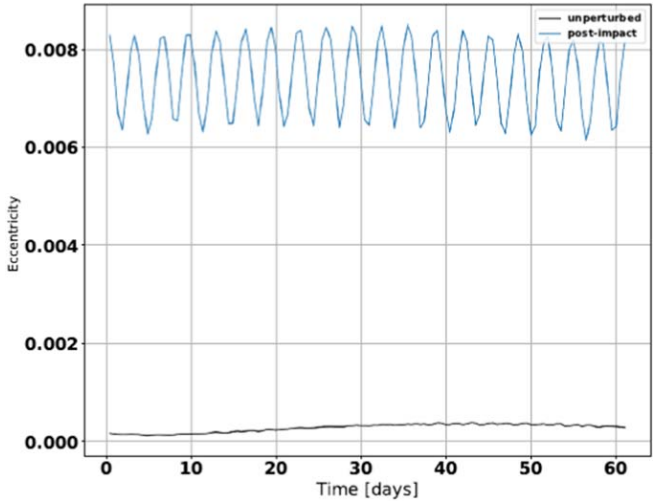

**Figure 4.** GUBAS simulation results showing the orbit eccentricity behavior over 60 days post-impact ($\beta = 1$), along with what the unperturbed behavior would be with no impact over the same time span. Note that the initial dynamically relaxed orbit has a very small eccentricity, with small variations, whereas the post-impact orbit eccentricity is oscillatory but bounded over the duration shown here.

observations of the period change, and additional analysis would be needed for those cases. That said, the non-DRA-compliant shapes for Dimorphos that result in the higher post-impact eccentricities are less likely to be the actual shape of Dimorphos.

### 3.2. Induced Libration and Other Excited States

For a circular orbit, the free libration frequency of a satellite is given by Murray & Dermott (2000)

$$\omega_{\text{lib}} = n \left[ \frac{3(B - A)}{C} \right]^{1/2}, \quad (2)$$

where $n$ is the mean motion and $A$, $B$, and $C$ are the satellite's three principal moments of inertia ($C > B > A$). At present, the shape of Dimorphos (and, therefore, its moments of inertia) is poorly constrained, and will likely not be known to significantly higher precision until the DART spacecraft arrives. Since Dimorphos's spin dynamics depend strongly on its moments of inertia, this represents a major source of uncertainty when estimating the system's post-impact dynamical state.

Agrusa et al. (2021) studied Dimorphos's post-impact attitude stability as a function of target body shape and the momentum transfer enhancement factor, $\beta$. Dimorphos was assumed to be a uniform-density ellipsoid parameterized by the semi-axis ratios $a/b$ and $b/c$, where $a > b > c$. This parameterization over body shape was deliberately chosen, since the shape will be determined from DART and LICIACube images of Dimorphos, rather than measurements of inertia moments. However, there is a simple one-to-one correspondence between body shape and inertia for a homogeneous triaxial ellipsoid.[27] Agrusa et al. (2021) began by deriving an analytical model in which Didymos and Dimorphos are treated as a sphere and a triaxial ellipsoid, respectively, and the system is assumed to be at equilibrium (i.e., in a circular orbit with Dimorphos in synchronous rotation). Four fundamental frequencies of motion were identified over the parameter spaces of $1.0 \leqslant a/b \leqslant 1.5$ and $1.0 \leqslant b/c \leqslant 1.5$, corresponding to the mean motion and Dimorphos's free libration, precession, and nutation frequencies. The analysis showed that these frequencies can become commensurate at a multitude of locations throughout the axis ratio parameter space, and that chaotic motion or attitude instability could be possible near these resonances.

Agrusa et al. (2021) then simulated the mutual orbit with a numerical model that accounts for Didymos's $J_2$ moment, representing its shape oblateness, and that integrates the mutual orbit in 2D and Dimorphos's attitude in 3D. This confirmed that Dimorphos's spin state becomes highly excited near the resonance locations. In addition, a fast Lyapunov indicator

---

[27] For a uniform triaxial ellipsoid, the moments of inertia $A$, $B$, and $C$ are related to the respective semi-axis lengths $a$, $b$, and $c$ through the following relations: $A = \frac{m}{5}(b^2 + c^2)$, $B = \frac{m}{5}(a^2 + c^2)$, and $C = \frac{m}{5}(a^2 + b^2)$, where $m$ is the ellipsoid mass.

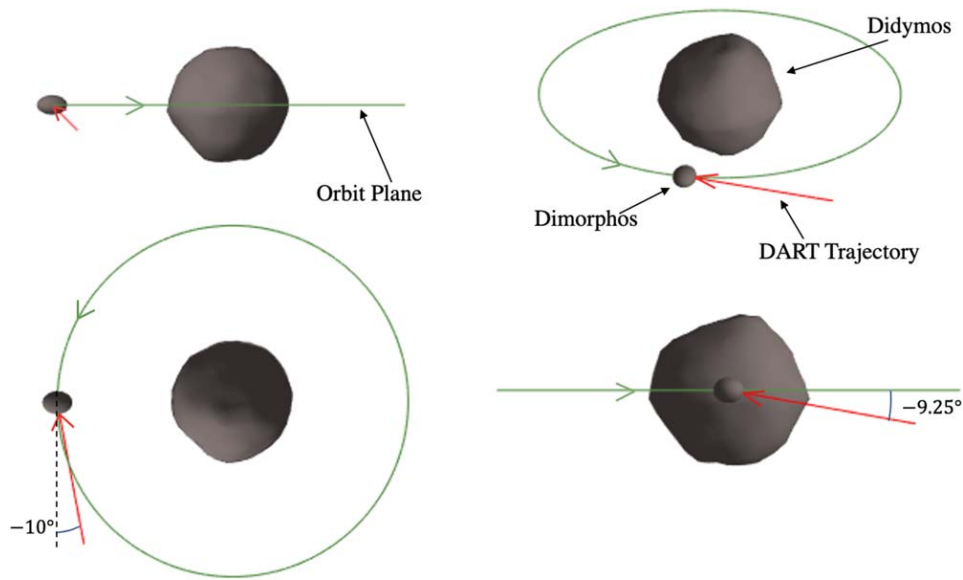

**Figure 5.** A schematic showing DART's expected impacting momentum vector, based on the latest flight trajectory consistent with the 2021 launch. The impact will be *nearly* head on, opposite the instantaneous orbital velocity of Dimorphos. In an RTN coordinate system, DART's momentum vector will have small components in both the radial and normal directions. DART will be coming from below the Didymos–Dimorphos mutual orbit plane, with an angle of $\sim-9.25°$ (the normal component), and it will also have an in-plane angle of $\sim-10°$ (the radial component).

analysis showed that the secondary's attitude evolved chaotically near resonance locations.

Finally, Agrusa et al. (2021) ran fully coupled F2BP GUBAS simulations over the parameter spaces of $1.0 \leqslant a/b \leqslant 1.5$ and $1.0 \leqslant b/c \leqslant 1.5$, with values for $\beta$ ranging from 1 to 5. In these simulations, the radar-derived polyhedral shape model was used for Didymos, while Dimorphos was treated as a uniform triaxial ellipsoid. The simulations were carried out for 1 yr of simulated time. The DART impact was assumed to be an idealized, head-on, planar impact, so momentum was transferred entirely within the mutual orbit plane and opposite to Dimorphos's instantaneous orbital velocity. These simulations agreed with the simpler models, showing that Dimorphos could become attitude-unstable near the resonance locations. The key difference between the GUBAS simulations and the simpler "$J_2$-ellipsoid" model was that GUBAS generally predicted lower post-impact libration amplitudes, which is a consequence of spin–orbit coupling. Since spin–orbit coupling is modeled self-consistently in GUBAS, there is a periodic exchange of angular momentum between Dimorphos's spin and the mutual orbit, which regulates Dimorphos's libration amplitude.

However, the real DART impact will not be an idealized head-on impact, and DART's momentum will have nonplanar components. To address this, the work of Agrusa et al. (2021) was extended to include a more realistic impact geometry. DART's current trajectory nominally results in an impact angle of $\sim-9.25°$ relative to the orbit plane, meaning that DART will have a small component of its momentum coming "up" from below the orbit plane.[28] For this trajectory, DART must also have an in-plane impact angle of $\sim-10°$ in order to improve the communication with ground stations.[29] Figure 5 shows the expected impact geometry. In the idealized head-on impact case, all momentum is transferred in the instantaneous tangential direction in a radial–tangential–normal (RTN) coordinate system sense. However, the realistic case will have small components of momentum transferred in the normal (due to the nonplanar impact angle) and radial (due to the in-plane impact angle) directions.

Figure 6 shows the maximum Euler angles (roll, pitch, and yaw) achieved by Dimorphos over a 1 yr simulation, for $\beta = 3$. Figure 6(a) shows the idealized planar case from Agrusa et al. (2021), while Figure 6(b) accounts for the current best estimate of the impact geometry. The resulting attitude evolution of Dimorphos is quite similar between the planar and nonplanar impact geometries. Dimorphos's attitude stability depends mainly on its own shape (i.e., moments of inertia) and the eccentricity of the mutual orbit. Since the geometry of the nonplanar impact is close to the idealized planar case, the resulting changes to the mutual orbit and attitude evolution match quite well. The most noticeable difference between the two cases is in the maximum roll-angle plot, where there are fewer cases in which Dimorphos enters a rolling state or "barrel instability" about its long axis (indicated by the roll angle exceeding 90°). This is due to the idealized planar impact being more efficient at changing the eccentricity and therefore increasing the range of Dimorphos shapes that become attitude-unstable. In the nonplanar impact, a small amount of momentum is "wasted" by being transferred in the normal and radial directions, rather than completely in the tangential direction.

### 3.3. Variations in the Orbit Period

The libration in Dimorphos following the DART impact will cause the mutual orbit period to fluctuate, as angular momentum is transferred between Dimorphos's rotation and the orbit. This nonconstant orbit period is discussed in detail by Meyer et al. (2021). Those results are expanded upon here by using fully coupled GUBAS simulations and the actual 3D impact geometry for an ellipsoidal secondary with axis ratios $a/b = 1.3$ and $b/c = 1.2$ (the nominal values from the DRA). The primary is modeled using the polyhedron radar shape model from the DRA.

---

[28] In fact, the binary's obliquity to its heliocentric orbit is nearly 180°, but for ease of discussion, the two-body angular momentum is taken to be oriented "upward," both here and in Figure 5.
[29] This means that DART will impact ~20 minutes before the moment when the impact would have been head on. In other words, a small amount of momentum will also be transferred in the instantaneous radial direction.

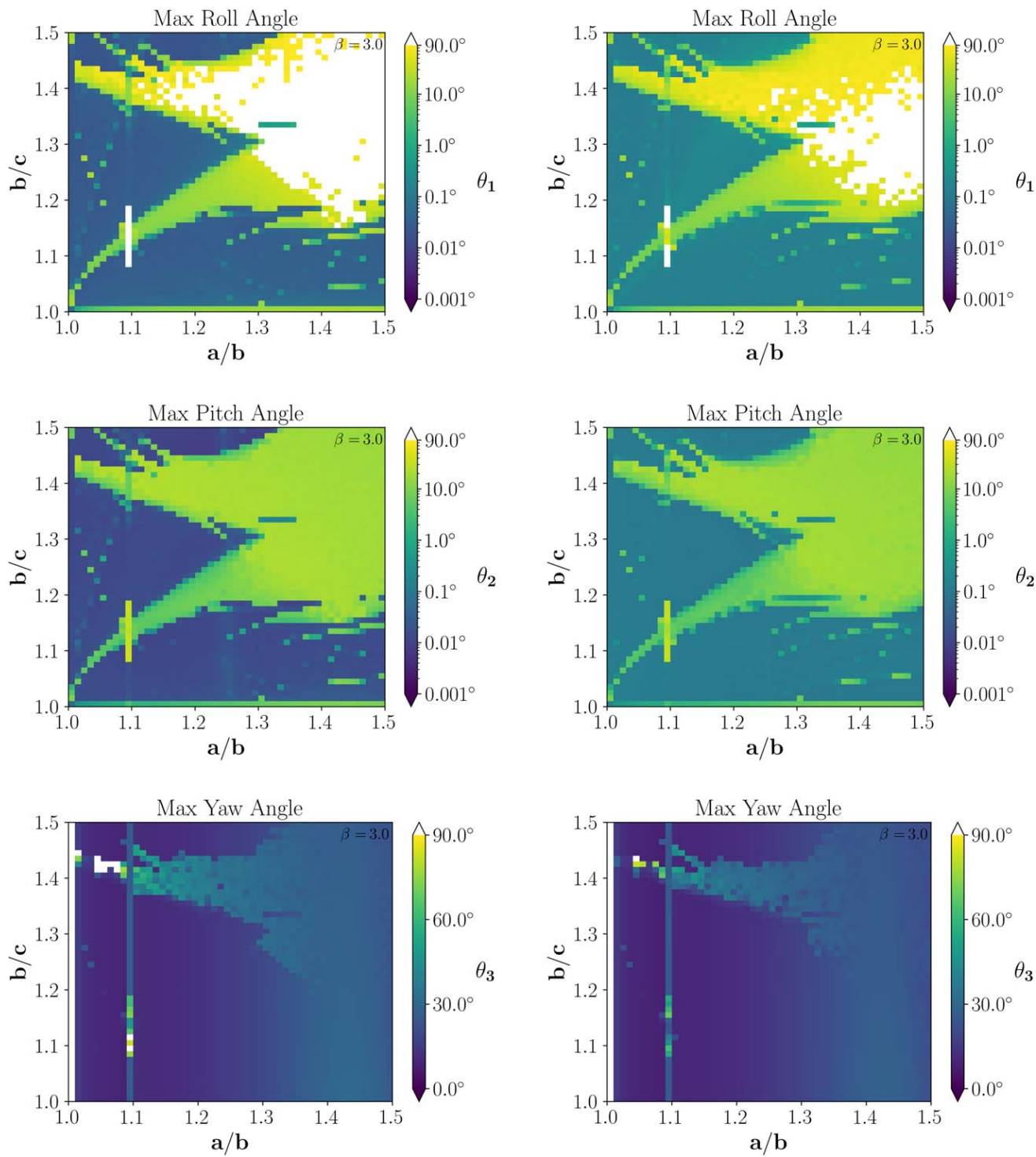

**Figure 6.** Comparisons of the attitude of Dimorphos resulting from (a) an idealized planar impact versus (b) the expected impact geometry for the current DART trajectory. Since the nonplanar components of DART's velocity vector are relatively small, the realistic impact geometry closely matches the idealized planar case.

Generally, the orbit period variations are driven by short-period and long-period modes. The short-period oscillations repeat once every several days, while the long period repeats on the order of several months. Both modes carry significant amplitudes, ranging from tens of seconds to a few minutes. Meyer et al. (2021) point out that the orbit period variations share the same frequencies as the eccentricity vector oscillations and the apsidal precession of the Keplerian orbit. The libration amplitude and variation in the orbit period are shown in Figure 7 as a function of $\beta$ for the 3D impact geometry. There is a strong linear trend for small values of $\beta$, but there is a notable increase in magnitude at $\beta = 5$. This is due to the impact imparting enough momentum to cause Dimorphos to begin spinning about its long axis, the so-called barrel instability first identified by Cuk et al. (2020), and further discussed by Meyer & Scheeres (2021), Agrusa et al. (2021), Quillen et al. (2022), and Ćuk et al. (2021). This phenomenon causes the orbit angular momentum to undergo larger fluctuations, and is absent at smaller values of $\beta$. Interestingly, the large increase in magnitude at $\beta = 5$ is absent in the 3D analysis of Meyer et al. (2021), using an asymmetric shape model for Dimorphos, again demonstrating the importance of Dimorphos's shape in the resulting dynamics (in the present analysis, an ellipsoidal Dimorphos was used). The out-of-plane rotation destroys the two-mode structure of the orbit period variations in favor of more chaotic variations, consistent with the findings in Meyer et al. (2021). Therefore, while fluctuations in the mutual orbit period are expected from the DART impact, if enough momentum is imparted to cause Dimorphos to begin tumbling, the orbit period fluctuations may be even larger than originally predicted.

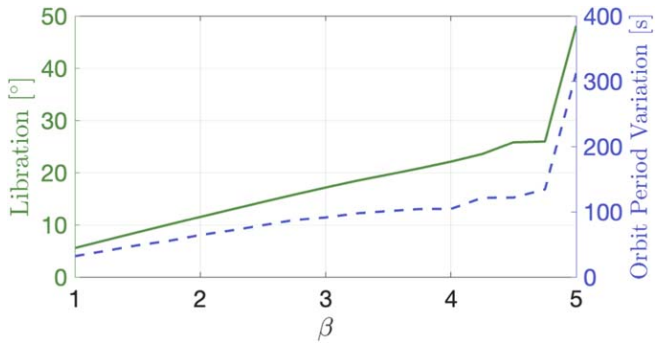

**Figure 7.** The libration amplitude and orbit period variation for Didymos using the full 3D impact geometry simulated in GUBAS for the case $a/b = 1.3$, $b/c = 1.2$. The libration amplitude, shown as the solid green curve, corresponds to the left axis while the orbit period variation, shown as the dashed blue curve, corresponds to the right axis.

While orbit period variations may pose a challenge for characterizing the post-impact system, they may also provide an opportunity. With a more accurate Dimorphos shape model obtained from DART and LICIACube images, observations of orbit period variations may help constrain the libration amplitude and possibly provide a supplemental estimate of $\beta$. This is left as future work following the impact.

### 3.4. Didymos System Heliocentric Orbit Changes

The DART impact will change the heliocentric orbit of the Didymos system by a small amount, which may be measurable through a combination of ground-based, DART, and Hera measurements. While measuring the change in Didymos' heliocentric orbit is not required to complete DART mission objectives, it can yield additional insights into the physical properties of the system. To quantify the heliocentric push, numerical simulations of the DART impact and the resulting ejecta were carried out in Stickle et al. (2020) and Fahnestock et al. (2022). The simulations track the states of the ejecta particles coming off Dimorphos as a result of the impact. That data is used to quantify the changes in the system's heliocentric orbit.

The momentum carried by the ejecta leaving the system contributes to the overall change in the heliocentric orbit. The cumulative momentum of all the particles that escape the Didymos system for the nominal F2BP case, normalized by the DART impact momentum, was calculated from GUBAS ejecta dynamics simulations. For this specific case, 95% of the ejecta mass escapes the system, carrying 88% of the ejecta momentum. The escape criterion for the particles is defined as the moment when they cross the system's Hill sphere, the approximate radius of which is

$$r_H = r \sqrt[3]{\frac{m}{3M_\odot}}, \quad (3)$$

where $r$ is the distance between the Sun and the Didymos barycenter at the time of impact, $m$ is the mass of the Didymos system, and $M_\odot$ is the mass of the Sun. The contribution of the ejecta to the overall momentum transport for a head-on impact can then be expressed as a corresponding heliocentric momentum enhancement parameter, $\beta_\odot$:

$$\beta_\odot = \frac{p_{\text{DART}} + p_{\text{ejecta}}}{p_{\text{DART}}} = 1 + \frac{p_{\text{ejecta}}}{p_{\text{DART}}} \approx 1.789, \quad (4)$$

where $p_{\text{DART}}$ is the momentum of the DART spacecraft at impact and $p_{\text{ejecta}}$ is the cumulative momentum of the simulated system-escaping ejecta particles at their corresponding time of escape. The corresponding heliocentric $\Delta v$ magnitude for the system in this case is about $1.06 \times 10^{-5}$ m s$^{-1}$. The value for this heliocentric $\beta_\odot$ is smaller than the $\beta$ value of 1.894 calculated for the particles that escape Dimorphos's gravitational pull, because not all momentum-carrying ejecta end up leaving the binary asteroid system's Hill sphere. The ejecta particles that remain in the system or get redeposited on either asteroid do not contribute to the heliocentric changes of the system. Makadia et al. (2022) concluded that the changes to the heliocentric orbit of Didymos imparted by DART are indeed minor and cannot send the Didymos system on a collision course with Earth after the DART impact.

The proportion of ejecta mass and momentum that ends up leaving the system is highly dependent on the surface strength and internal friction of Dimorphos. A measurement of both $\beta$ and $\beta_\odot$ could produce additional constraints on the internal structure of Dimorphos. By far, the greatest influence on both $\beta$ and $\beta_\odot$ is the target cohesion. For a strong (>10 kPa) target, a $\beta$ smaller than about 2.5 is expected (Raducan et al. 2019; Stickle et al. 2020). In these scenarios, most of the material is ejected out of the crater at speeds larger than the escape speed of the Didymos system, and most of the momentum leaves the system. On the other hand, recent studies (e.g., Raducan & Jutzi 2022) show that for targets with cohesion less than ≈10 Pa, more target mass is ejected at speeds lower than the escape speed of the system (and is therefore trapped in the system) than at speeds higher than the escape speed of the system. This happens because with decreasing cohesion, the cratering efficiency and the total mass of ejected material also increase. This means that for weak targets, more ejecta mass remains in the system than leaves the system. In these target scenarios, up to 20% of the momentum remains in the system.

Observations of the ejecta cloud combined with lightcurve measurements of the Didymos system right after impact will help determine the change in the mutual orbit period and the local system's $\beta$ value. Comparing independent estimates for the value for $\beta_\odot$ from changes in the heliocentric orbit to the local system's $\beta$ value would allow us to put additional constraints on the physical properties of Dimorphos, such as porosity and/or strength. However, such estimates will require separate ground- and/or space-based observations.

**Table 3**
Estimated Accuracy for Retrieved $\beta_\odot$ for Various Observation Campaigns[a]

| Observation Campaign | $\beta_\odot$ Estimate | $1\sigma$ Uncertainty |
| --- | --- | --- |
| 2 Hera | failed | failed |
| 12 Radar + 2 Hera | failed | failed |
| 2 Occultations + 2 Hera | 1.793 | 0.130 |
| 3 Occultations + 2 Hera | 1.798 | 0.104 |
| 12 Radar + 2 Occultations + 2 Hera | 1.798 | 0.104 |
| 12 Radar + 3 Occultations + 2 Hera | 1.796 | 0.102 |

**Note.**
[a] The correct value to be retrieved is $\beta_\odot = 1.789$.

To determine which observation strategy, if any, would result in accurate and well-constrained estimates for $\beta_\odot$, simulations of various combinations of observation techniques were carried out and analyzed to determine what effect they would have on the knowledge of Didymos's heliocentric orbit before and after the impact. Table 3 shows the $\beta_\odot$ estimate resulting from possible DART observation campaigns. Optical and radar astrometry for the Didymos system barycenter was simulated by propagating the nominal orbit of the system beyond the epoch of impact. The change in asteroid momentum was taken into account and modeled according to the $\beta_\odot$ value given by Equation (4). Each of these simulated observations were given expected accuracies for the Hera range measurements (1 m), the occultation angular measurements (0.5 mas), and the radar-based range measurements (15 m; Naidu et al. 2020). These observations were then passed through a least-squares orbit determination and parameter estimation algorithm to recover the value of $\beta_\odot$. All runs included not only the simulated observations, but also the actual Didymos observation data available from the IAU Minor Planet Center as of 2022 January. In Table 3, "Occultations" refer to a series of high-accuracy stellar-occultation observation opportunities. The three occultation scenarios consist of one before the DART impact, in 2022 September, and two after the impact, in 2022 November and 2023 January, whereas the two occultation cases only consider the two opportunities after the DART impact. Currently, 12 radar observations are expected—three before and nine after the impact, with an interval of 1 day between them. It appears that occultation observations are likely necessary to estimate the change in the heliocentric orbit of the Didymos system effected by DART. Although determining the heliocentric $\beta$ is not part of the DART mission's primary requirements, it may be feasible and, along with the local $\beta$ measurement, it can help further constrain the physical properties of Dimorphos.

### 3.5. The Instantaneity Assumption

The cratering process on a large body is typically fast in relation to the dynamics of its system. On a small body like Dimorphos, with low strength and low gravity, and as part of a binary, the finite time that it takes to form a crater and produce ejecta can become significant in comparison to its dynamical environment. A complete and precise determination of $\beta$, the ejecta evolution, and thus the resulting dynamical state of the Didymos system, should ideally take into account both the time evolution of ejecta production and the displacement of the launch position from the impact point. The Hayabusa2 sample-return mission to 162173 Ryugu (Tsuda et al. 2013) included the SCI experiment, which consisted of a 2 kg copper projectile impacting the surface at a velocity of 2 km s$^{-1}$ near-normal to the surface (Saiki et al. 2017). Images from the Hayabusa2 DCAM3 imager revealed an ejecta blanket still attached to the surface (and still producing ejecta) after ~500 s and a crater about 15 m in diameter (Arakawa et al. 2020). Dimorphos is significantly smaller than Ryugu, and has a different composition, and the DART impact will be ~3 orders of magnitude greater in energy than the SCI impact (Arakawa et al. 2017). In all, it is expected that the crater will be larger on Dimorphos and that the cratering event will last longer (Stickle et al. 2022). To date, most ejecta simulations predicting the DART impact have assumed that all ejecta are produced and launched at the same point in time and space (e.g., Yu et al. 2017; Yu & Michel 2018; Fahnestock et al. 2022).

Although the momentum transfer and cratering process takes a finite amount of time, post-impact dynamics simulations of the Didymos system to date have treated the DART impact as an instantaneous perturbation to Dimorphos's velocity and rotation state (Section 3.1). In addition, the mass of Dimorphos is unchanged in dynamics simulations, which ignores mass loss due to crater formation and ejecta production. As long as the crater size and ejected mass are small compared to Dimorphos and the momentum transfer time is significantly less than the system's orbit period of ~12 hr, this latter approximation is valid. Recent studies have shown that if Dimorphos has a cohesive strength $\gtrsim 10$ Pa, the crater size and ejected mass should be small compared to Dimorphos's diameter and mass (Stickle et al. 2017; Raducan et al. 2019). However, if Dimorphos is a weak target, non-negligible shape changes may occur; this possibility is discussed in further detail in Section 5 (Raducan & Jutzi 2022). To account for finite crater formation and ejecta production times, momentum transfer processes that last up to 30 minutes have been tested in GUBAS F2BP simulations, assuming a center-of-mass impact, an impact vector aligned with the surface normal, and ejecta that readily leave the system. These simulations show a negligible difference between instant and time-dependent momentum transfer, which implies, given these assumptions, that the system dynamics (i.e., the change in orbital period) is not sensitive to the ejecta production time, per se, for reasonable transient crater formation times. If the important contributions to $\beta$ are made by ejecta that readily leave the system, then DART's Level 1 requirements can be achieved by assuming instant momentum transfer and inferring the "total" $\beta$ from a measured change in the orbital period, regardless of how the contributions to $\beta$ are partitioned in time among ejecta particles. However, if the target material conditions are such that a substantial amount of momentum will be contained in the slow-moving ejecta (relative to the gravity of the Didymos system) produced in the late stages of crater formation, then a more careful consideration of the contribution to $\beta$ is required, one that specifically treats the production and trajectories of slow-moving ejecta.

An ejected particle's contribution to $\beta$ is not over once it leaves Dimorphos's surface. For example, if it reimpacts nearby soon after, with opposite velocity, its momentum contribution is zero. However, even if the particle does not reimpact, it continues to contribute gravitational work until it escapes the system. Considering this effect, Holsapple & Housen (2012) apply a first-order analytical correction to $\beta$ by considering the change in velocity of an ejected mass element from when it leaves the surface to when it reaches infinity, assuming a kinetic impact into a singleton asteroid, whose cratering process is dominated by a power-law profile and ignoring solar tides (Jutzi & Michel 2014 also make use of this correction when estimating $\beta$ from their smoothed-particle hydrodynamics simulations). This accumulated change in ejecta velocity after launch is most significant for low-speed ejecta, as they take longer to leave the gravitational influence of the target.

Ejection velocities steadily decrease as the transient crater expands. For much of its excavation, the crater ejects material with a velocity profile described by a power-law flow. At later stages, this power-law relationship breaks down. The material properties of the target and the speed of the launching ejecta at this point in time, $v_{\text{break}}$, are enough to provide an effective description of the velocity profile of the ejecta, from which the analytical correction to $\beta$ can be derived. When the strength of the target is sufficiently low, the power-law flow during the transient crater excavation can proceed until the ejecta speeds are below the target's escape speed, $v_{\text{esc}}$, increasing the overestimate of $\beta$ when the assumption of instantaneous momentum transfer is used. For porous granular material, like sand grain packing on Earth, the Holsapple & Housen (2012) correction implies decreases in $\beta$ of about 7.5%, 19.5%, and 33% for $v_{\text{break}}/v_{\text{esc}}$ of 1, 2, and 5, respectively. For still higher porosity, however, it is expected that a larger portion of the ejecta would be launched at faster speeds (Housen & Holsapple 2011), decreasing these estimates and reducing the difference in $\beta$.

For DART, in the case of a rubble-pile Dimorphos, $v_{\text{break}}$ may be comparable to or exceed $v_{\text{esc}}$, which makes understanding the ejecta profile and the finite time span of the crater formation an important topic of ongoing study. There is also the added complication of Didymos's gravitational influence (the Dimorphos Hill sphere, Equation (3), with respect to Didymos, is only about twice the size of Dimorphos itself). In addition, the fact that the binary orbital phase will change during the evolution of low-speed ejecta further complicates the situation.

One approach to going beyond idealized analytical estimates to account for the temporal variation in ejecta production as well as the spatial extent of the launch position is to take the outputs from hydrodynamic impact codes as the inputs—or initial conditions—of a granular dynamics code for computing all but the earliest stages of the ejecta production, evolution, and the formation of the transient crater. The DART Science Team is at work on producing such a procedure; preliminary efforts include Schwartz et al. (2016), Zhang et al. (2021), and Ferrari et al. (2022). The DART mission provides a unique opportunity to understand the time-dependent nature of the cratering and momentum transfer process.

## 4. Contributions of Internal Structure to Dynamics

Rigid-body simulations (Agrusa et al. 2021, and see Section 3) show that, depending on its shape/inertia moments, Dimorphos can become attitude-unstable and its spin state may evolve chaotically as a result of the DART impact. However, according to the current understanding of binary formation and evolution history, Dimorphos may be a rubble pile (Richardson et al. 2002; Walsh et al. 2008; Walsh & Jacobson 2015; Tardivel et al. 2018). In this context, the nonrigid, deformable internal structure of Dimorphos may alter its dynamical

Table 4
Range of Parameters Considered in Rubble-pile Simulations around the Nominal Values in Table 1

| Body | Parameter | Value Range |
|---|---|---|
| Didymos | Mass (kg) | $(5.350\text{–}5.355) \times 10^{11}$ |
|  | Bulk density (kg m$^{-3}$) | 2155–2182 |
|  | Rotation period (hr) | 2.26 |
|  | Shape | Point mass |
| Dimorphos | Mass (kg) | $(4.887\text{–}4.923) \times 10^{9}$ |
|  | Bulk density (kg m$^{-3}$) | 2155–2182 |
|  | Rotation period (hr) | 11.921 7 |
|  | Shape | Triaxial ellipsoid |

**Note.** The shape and bulk density of the rubble-pile model are evaluated by using its DEEVE, an ellipsoid with a uniform density and the same volume and moments of inertia as the rubble-pile aggregate.

response to DART's impact, compared to the rigid-body case. To assess whether this makes an important difference on the post-impact dynamics, two cases were selected from the study by Agrusa et al. (2021) and replicated with an $N$-body discrete element method (DEM) code: one case where Dimorphos becomes attitude-unstable in the rigid-body simulations, and one where it remains in a stable libration state. Since rubble-pile simulations are far more computationally expensive than rigid-body simulations, they are necessarily more limited in scope. This section focuses mainly on changes to the rotational motion of Dimorphos and its attitude stability that result from a change in the mutual orbit (Section 5 discusses shape changes to Didymos and Dimorphos that could result directly from the DART impact). The results are summarized below; we refer the reader to Agrusa et al. (2022) for further details.

Overall, the results show that a DEM model of Dimorphos does not appreciably alter its stability properties for the cases studied thus far. This implies that, for the same shape of Dimorphos, rigid F2BP simulations should largely agree with corresponding higher-fidelity DEM models, with minor disagreements attributable to small (and unavoidable) differences among the simulation setups and the numerical routines of the respective codes.

### 4.1. The Rubble-pile Model

The dynamics of the Didymos binary system were studied with rubble-pile, self-gravitating models of the system components, using the $N$-body DEM code PKDGRAV (Richardson et al. 2000; Stadel 2001). Contact interactions between particles are handled using the soft-sphere discrete element method (SSDEM; Schwartz et al. 2012). PKDGRAV has already been used and validated in rigid F2BP studies of the Didymos system (Agrusa et al. 2020); here, the analysis is extended by enabling the code's SSDEM option to model Dimorphos as a rubble pile, rather than a rigid aggregate body.

Table 4 summarizes the parameter ranges considered for the Didymos system for this study. Guided by Agrusa et al. (2021), two simulation cases were considered, representing two possible shapes of Dimorphos. Dimorphos is assumed to be a triaxial ellipsoid (Naidu et al. 2020), with semi-axis ($a > b > c$) ratios and values as follows:

1. *Stable* case: $a/b = 1.2$, $b/c = 1.1$; (95.00, 79.17, 71.97) m.

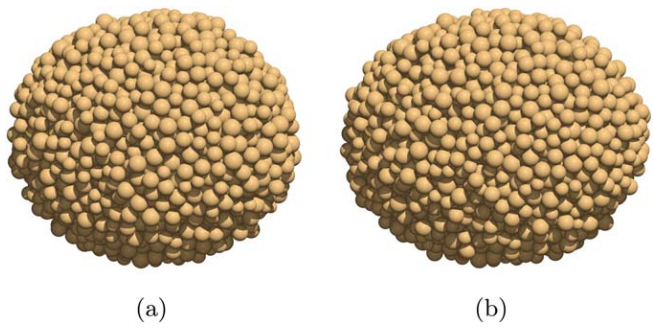

**Figure 8.** Top-down views of the rubble-pile models of Dimorphos in PKDGRAV. (a) is the stable case, with $a/b = 1.2$ and $b/c = 1.1$, while (b) is the unstable case, with $a/b = 1.3$ and $b/c = 1.2$. Despite the small apparent difference in their respective shapes, the resulting spin evolutions of these two rubble piles will differ significantly. Didymos is not included in this visualization, since it is just a point mass in these simulations.

2. *Unstable* case: $a/b = 1.3$, $b/c = 1.2$; (103.16, 79.35, 66.13) m.

The rubble-pile models are built to closely match the semi-axis lengths and moments of inertia of the equivalent rigid-body cases, using ~4000 randomly packed spherical particles. The physical properties and overall shape of the rubble-pile model are evaluated using its dynamically equivalent equal-volume ellipsoid (DEEVE), and are set to match a uniform-density ellipsoid, with the same volume and inertia as the ideal, rigid, triaxial ellipsoid. Due to its discrete nature, the inertial and geometrical properties of the rubble-pile model can never exactly match those of an ideal, homogeneous ellipsoid. The external surface of a rubble-pile model is not precisely defined and has uncertainties on the order of the size of a single fragment, which, in this case, is a few meters, compared to the overall ~200 m long ellipsoid. Although this discrepancy is small, it is enough to perturb the rigid-body equilibrium condition, which is not reproduced exactly. This structural limitation due to DEM discretization, alongside the high sensitivity of the problem to initial conditions, challenges the precise reconstruction of the dynamical state of the system. However, this is shown not to affect the ultimate fate of the simulation, i.e., whether the system evolves toward a stable, bounded behavior or an unstable behavior.

Both the unperturbed dynamics of Didymos, with asteroids on a mutual circular orbit about the barycenter of the system, and the post-impact case, where the orbital velocity of Dimorphos is instantaneously perturbed as a proxy for a $\beta = 3$ DART-like impact, were investigated. In-plane motion and no radial component of orbital velocity at the initial time are assumed for both Didymos and Dimorphos. Compared to the rigid case, the orbital initialization of the rubble-pile case is not straightforward, as the rubble pile suddenly feels a tidal stress that causes it to deform slightly. To address this issue, Dimorphos is first simulated as a rubble pile in orbit around Didymos for ~2 orbital periods, to allow Dimorphos's shape to settle into equilibrium before the DART perturbation is applied.

Figures 8(a) and (b) show a top-down view of the two rubble-pile models of Dimorphos. The stable and unstable shapes look relatively similar, with their longest axes differing by ~15 m relative to their total length of ~200 m. This small difference in shape (and therefore moments of inertia) nonetheless leads to a significant difference in their respective attitude evolutions, as found for the rigid models.

### 4.2. Spin Motion of Dimorphos

The spin motion of Dimorphos can be quantified by studying the time evolution of Dimorphos's three Euler angles that describe its attitude: roll, pitch, and yaw. In the idealized, tidally locked equilibrium configuration, these angles are zero. But since the discrete model cannot reproduce exactly the same shape and axis ratios of an ideal rigid body, the system achieves at best a nonperfect, tidally locked equilibrium, even in the stable case ($a/b = 1.2$, $b/c = 1.1$). In this case, the amplitude of the libration is not zero, but always remains bounded and never becomes unstable.

Figures 9(a) and (b) show the long-term evolution of Dimorphos's attitude, with a comparison between the GUBAS rigid-body case and the PKDGRAV rubble-pile case. The difference between the "rigid" and SSDEM PKDGRAV cases is that the particles making Dimorphos are locked into a rigid aggregate, meaning that it has *nearly* the exact same shape and moments as the SSDEM case, making it useful for direct comparisons. Figure 9(a) shows the stable case, when Dimorphos has a shape of $a/b = 1.2$ and $b/c = 1.1$. This plot is consistent with the Agrusa et al. (2021) prediction when Dimorphos is modeled both as a rigid body and as a rubble pile (with the SSDEM option) in PKDGRAV. The rigid and SSDEM cases show very similar behaviors, with the roll and pitch angles remaining near zero as the body steadily librates (yaw motion) with an amplitude of ~20°. However, when an expected attitude-unstable case is simulated, the rigid and SSDEM models begin to diverge. Figure 9(b) shows Dimorphos's attitude when the body shape is changed to $a/b = 1.3$ and $b/c = 1.2$. All three simulations show Dimorphos becoming attitude-unstable. However, both of the PKDGRAV simulations (both rigid and SSDEM) become unstable before the equivalent GUBAS simulation. This is likely due to a combination of the two codes using different numerical integrators and Dimorphos's shape in PKDGRAV differing slightly from the idealized GUBAS case, as discussed previously.

Qualitatively, the two codes show broad agreement, indicating that the faster GUBAS code is adequate for capturing the system dynamics on timescales relevant for the DART mission. However, it may be possible that Dimorphos could reshape, either due to the DART impact directly or due to longer-term dynamical effects, in which case a DEM code like PKDGRAV would be necessary. A forthcoming detailed rubble-pile-focused study (H. Argusa 2022, in preparation) simulates Dimorphos with PKDGRAV under a wider range of body shapes, material parameters, and $\beta$ values. This study also includes simulations where Didymos is treated as a rubble pile rather than a point mass, and simulations using GRAINS (Ferrari et al. 2017, 2020), a DEM N-body code that allows for the use of nonspherical particles.

Furthermore, it is worth highlighting that a nonzero libration results in a change of the spin rate of Dimorphos, and, by conservation of angular momentum, any change in the spin rate of Dimorphos produces a change in its orbital motion around the barycenter of the system. This effect has been quantified for the rigid-body Dimorphos case by Meyer et al. (2021). The periodic variations closely follow the libration oscillation and are due to the instantaneous change of the asteroid's spin motion. In principle, for a rubble-pile object, secular variations might appear over long timescales as a result of the intrinsic dissipation between the rubble-pile constituents—friction between the aggregate building blocks—driven by tidal stresses (Nimmo & Matsuyama 2019). However, the extremely long timescale of this phenomenon is such that it is not practical for it to be studied using DEM models. A more detailed discussion on secular effects is provided in Section 6.

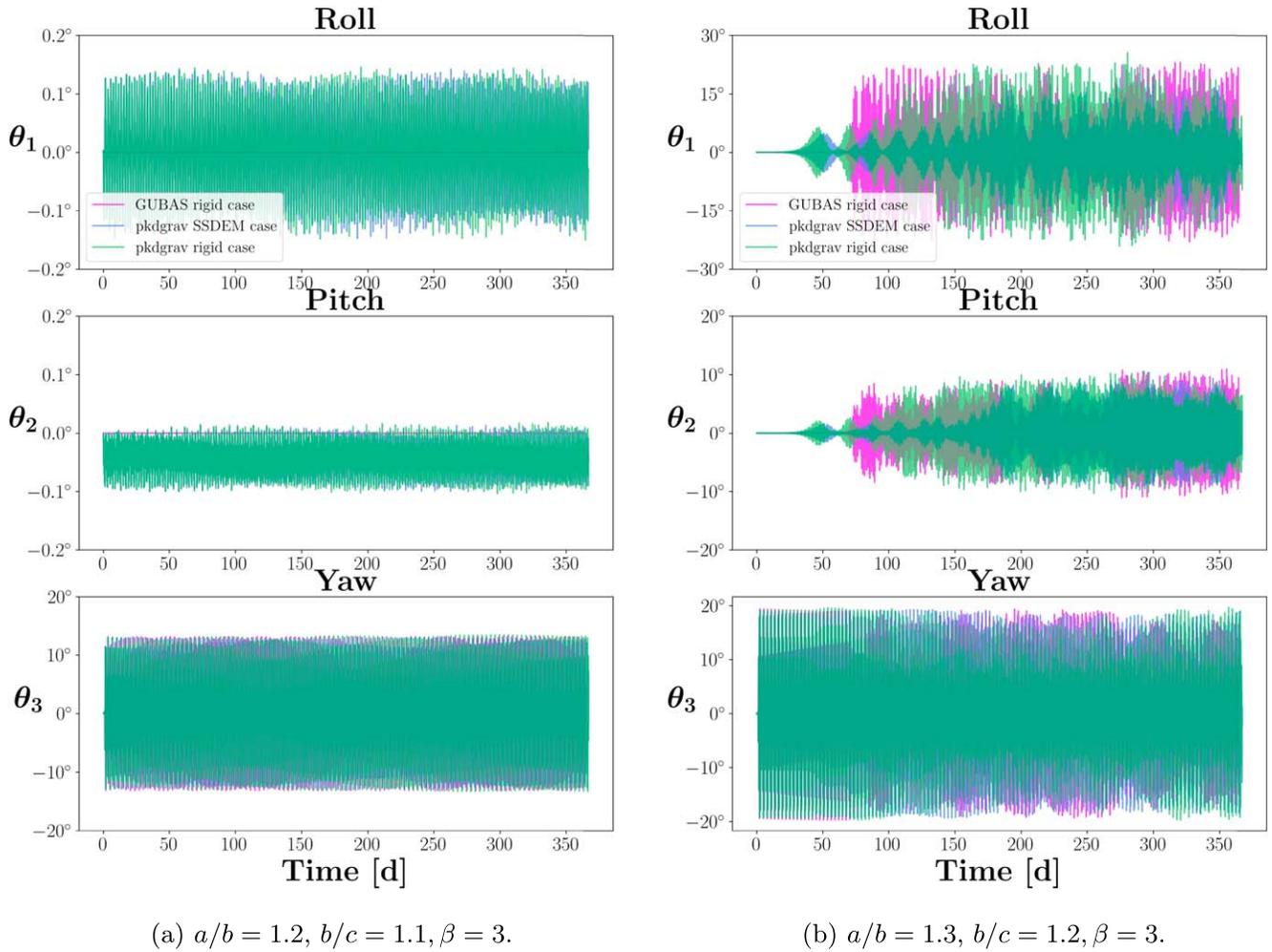

(a) $a/b = 1.2, b/c = 1.1, \beta = 3$.    (b) $a/b = 1.3, b/c = 1.2, \beta = 3$.

**Figure 9.** Time-series plots of Dimorphos's Euler angles (roll, pitch, and yaw) in the rotating orbital frame for the two body shapes under consideration. Figure 9(a) shows the expected stable case, which indeed remains stable when simulated both as a rubble pile (the SSDEM case) and as a rigid body, matching the GUBAS prediction. The only difference is that there are larger (but still small) oscillations in the roll and pitch angles in the PKDGRAV cases, which is likely due to the slightly different body shapes (and therefore different inertia moments) compared to the idealized ellipsoid used in the GUBAS simulation. Figure 9(b) shows the predicted unstable case, where both PKDGRAV cases show general agreement. The main difference between the two PKDGRAV cases and the GUBAS case is that the PKDGRAV cases become attitude-unstable sooner, likely owing to a combination of slightly different shapes compared to the idealized ellipsoid and a different numerical integrator.

## 5. Implications of Possible Didymos and Dimorphos Shape Changes

Section 4 shows that dynamical perturbations due to continuous deformation driven by mutual tides may negligibly affect the dynamical evolution over a short time period. In this case, the rigid-body assumption may suffice to characterize the system's dynamical behavior. On the other hand, the coupling behavior between deformation and dynamics may be enhanced, to cause energy dissipation due to tidal interactions between Didymos and Dimorphos. Such mutual tide-driven interactions, as well as other nongravitational effects, significantly influence the mutual orbit evolution (Section 6). In addition to these processes, there is a possibility that the instantaneous deformation processes occurring in either Didymos or Dimorphos will permanently change the mutual dynamics, affecting the inferred $\beta$ value.

Potential causes of reshaping include the following. First, Dimorphos experiences the DART impact, which generates a crater on its surface. Depending on the target surface and the impact conditions, the shape change from crater formation may perturb the mutual dynamics sufficiently to be detected by telescopic observations. This perturbation process results from changes in the mutual gravity interaction due to the reshaping. Second, Didymos is currently rotating at a spin period of 2.26 hr, while its bulk density is estimated to be $2170 \pm 350 \text{ kg m}^{-3}$ (Table 1). This condition implies that the asteroid's rotation is close to its critical spin limit. Thus, loose particles at the equator would barely remain on the surface, implying that Didymos may at present be sensitive to structural failure (Zhang et al. 2017; Naidu et al. 2020; Zhang et al. 2021). This suggests that there is a chance that Didymos may experience non-negligible reshaping as a consequence of the DART impact event (Hirabayashi et al. 2017, 2019), triggered when DART-driven ejecta (e.g., Richardson & O'Brien 2016; Yu et al. 2017; Yu & Michel 2018; Cheng et al. 2020; Raducan & Jutzi 2022) impart energy to Didymos's surface on contact. This reshaping process may also change the mutual dynamics.

Early numerical work has revealed the possibility of such scenarios, although, given the uncertainties of the geophysical conditions, whether such an incident is likely to occur and how it is likely to affect the mutual dynamics are poorly constrained. This section briefly summarizes recent efforts to investigate this issue, the details of which are given by Hirabayashi et al. (2022) and Nakano et al. (2022). The discussions below consist of two parts: the first part concerns the DART impact's influence on reshaping Dimorphos and any resulting dynamical changes, while the second explores Didymos's sensitivity to structural failure, along with any changes to the mutual orbit that result from a reshaped Didymos.

### 5.1. Dimorphos

#### 5.1.1. Influence of the DART Impact on the Structure

The kinetic energy imparted by the DART impact will affect Dimorphos's morphology locally or even globally, depending on the impact and target surface conditions. Supposing that the DART spacecraft (∼563 kg at impact) is an ideal sphere, with a radius of 0.41 m, and that it approaches Dimorphos with a speed of ∼6.14 km s$^{-1}$ at an impact angle of ∼−9.25° (Cheng et al. 2020), the $\pi$-scaling relationship with dry sand parameters (e.g., Holsapple 1993; Richardson 2009) predicts that the crater size is ∼32 m in diameter when the surface strength, $Y$, is 1 kPa (∼100 m when $Y = 0$ Pa). Raducan & Jutzi (2022) show that if Dimorphos's surface has a cohesive strength in excess of 10 Pa, the bulk of the excavated materials may have ejection speeds greater than the escape speed from the surface of Dimorphos, i.e., ∼9 cm s$^{-1}$. On the other hand, if the cohesive strength is below 10 Pa, the impact process is likely in the subcatastrophic regime, with less fast-particle ejection. Still, materials inside the body are significantly redistributed for this subcatastrophic case, resulting in non-negligible reshaping of the target body. These findings suggest that the reshaping magnitude could be constrained by characterizing the behavior of the ejecta launched from the impact site (mass, velocity field, and geometry).

#### 5.1.2. Reshaping-induced Mutual Orbit Perturbation

The reshaping of Dimorphos may induce perturbations to the mutual orbit. To investigate this possibility, a finite element method (FEM) approach was used to simulate the mutual dynamics of the Didymos system, by accounting for the reshaping of Dimorphos during the DART impact (Nakano et al. 2022). The model uses the shape models from the Didymos system DRA data (Section 2). The simulations assume that the Dimorphos shape change that is driven by the DART impact occurs on Dimorphos's leading side,[30] without considering the out-of-plane velocity component of the DART impact, given that such a component does not contribute to the in-plane motion significantly (Section 3.2). Also, each reshaping is modeled under the assumption that the volume and density distributions are constant before and after the DART impact (i.e., the ejecta volume is negligible compared to the volume of Dimorphos)—see Figure 10(a). The results show that the change to the mutual orbit period is linearly proportional to the reshaping magnitude (Figure 10(b)). For example, when the shape changes 8 m along the intermediate axis (see $L_0 - L$ in Figure 10(a)), the mutual orbit period, which is currently 11.9216 hr (Table 1), may become about 30 s shorter. On the other hand, if Dimorphos experiences a 4 m change along the same axis, the orbital period becomes 15 s shorter. From these conditions, because the observed orbital period change is linear, the derived relationship predicts that if the reshaping is larger than 2 m along the intermediate axis, the resulting orbital period change becomes larger than the DART measurement requirement, which is 7.3 s (Rivkin et al. 2021).

### 5.2. Didymos

#### 5.2.1. Structural Stability and Possible Internal Structure

As detailed above, Didymos, with its current spin and estimated bulk density, is close to its stability limit, unless it has mechanical strength or structural layers keeping the interior and surface structurally static. Recent efforts using DEM codes (Zhang et al. 2017, 2021; Ferrari & Tanga 2022) have shown two possibilities for Didymos's interior structure providing structural stability at fast spin rates: core mechanical strength or global weak cohesion.

Zhang et al. (2017) show that, with a bulk density close to the upper limit of the observed range, i.e., ⩾2380 kg m$^{-3}$, Didymos's structure can remain intact if the asteroid has the mechanical strength of a simulated crystallized structure with high friction. Ferrari & Tanga (2022) estimate that a very large rigid-core fraction (>50%) can keep the current structural configuration stable, with its nominal bulk density at the current fast spin, even when the external layer is made of cohesionless discrete materials. The rigid core does not have to be a single monolith, but it should be supported by some mechanical strength. The granular medium that constitutes the external layer can attain high angles of friction due to geometrical interlocking, as demonstrated in DEM simulations using angular particles (Hirabayashi et al. 2015; Sánchez & Scheeres 2018; Ferrari & Tanga 2020).

Alternatively, if there exists a substantial amount of dust or small grains in Didymos, the van der Waals forces between them could be strong enough to "cement" large particles and boulders, supplying adequate cohesive strength to discrete elements (Scheeres et al. 2010; Sánchez & Scheeres 2014). If this is the case, a rubble-pile structure with a low-strength interior could also remain intact. Recent studies predict that at Didymos's present spin, and with the nominal bulk density, cohesive strength higher than 10–20 Pa can enable Didymos's structure to remain intact (Naidu et al. 2020; Zhang et al. 2021). This prediction implies the existence of many fine grains with sizes of ∼2–5 $\mu$m (Sánchez & Scheeres 2014, 2016 note that this condition was calculated assuming spherical particles and a Hamaker constant derived for lunar regolith).

#### 5.2.2. Structural Sensitivity and Possible Failure Behaviors

If Didymos's structural properties are close to the stability requirement discussed above, Didymos would fail structurally when subject to an external destabilizing effect, e.g., YORP-induced spin-up and/or meteoroid impacts. There are likely three failure modes: internal structural deformation, fission of an asteroid body into components, and surface-mass shedding, where the latter two modes could be at the origin of the formation of Dimorphos (Walsh et al. 2008; Tardivel et al. 2018). The fission mode results from global crack propagation in the body, inducing body separation, while cohesion still binds grains together to form smaller coherent pieces. The presence of boulders up to the size of Dimorphos on the asteroid Ryugu (Watanabe et al. 2019) suggests that if such large boulders exist at Didymos's equator, they could detach from the surface in the absence of cohesion (Campo Bagatin et al. 2020). The mass-shedding mode may occur when Didymos has a relatively strong or denser core (Hirabayashi et al. 2015; Zhang et al. 2017; Sánchez & Scheeres 2018; Ferrari & Tanga 2022) or has a bulk density ≳2170 kg m$^{-3}$ with moderate cohesion (Zhang et al. 2021). Figure 11 shows one case with the internal deformation and mass-shedding failure modes, where Didymos is modeled as a polydisperse rubble pile with a bulk density of ∼2170 kg m$^{-3}$ and interparticle cohesion of ∼320 Pa (Zhang et al. 2021).

---

[30] This condition assumes that the system is in its dynamically relaxed state and that Dimorphos is in synchronous rotation (Section 2). Thus, the leading direction corresponds to the intermediate principal axis.

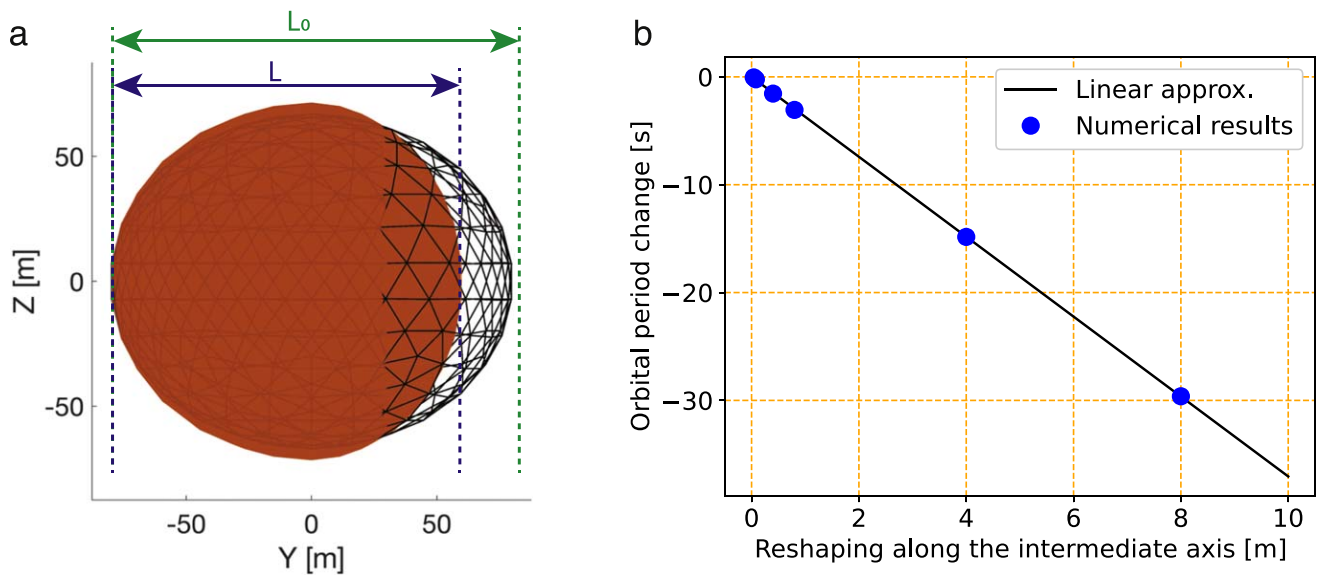

**Figure 10.** Dimorphos's deformed shape and the resulting orbit period change as a function of the reshaping magnitude. (a) An example of the reshaping, where the horizontal axis (the y direction along the intermediate principal axis, in contrast to the z-axis along the maximum principal axis) corresponds to the leading side. (b) Orbital period change as a function of deformation along the leading direction. This change is in addition to any change in Dimorphos's momentum due to the impact itself. The blue dots are derived from the FEM approach and the black line gives a linear approximation.

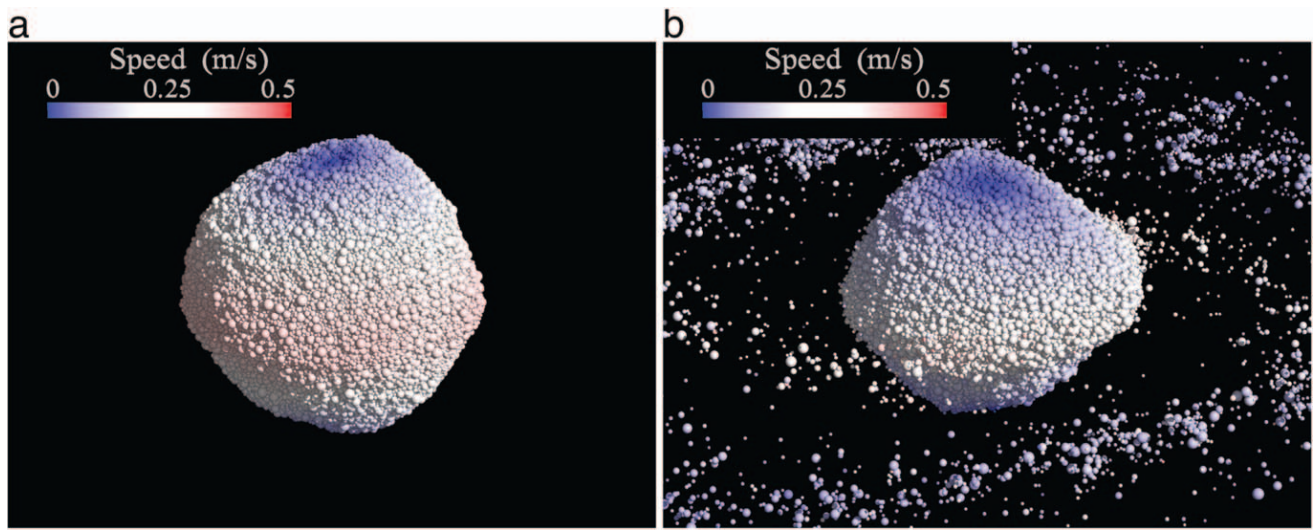

**Figure 11.** Didymos's deformation process, based on DEM simulations, where the interior is modeled as a dense polydisperse packing of grains with a particle diameter distribution ranging between ∼4 m and ∼16 m. The bulk density is ∼2170 kg m$^{-3}$ and the interparticle strength is ∼320 Pa (Zhang et al. 2021). The colors show the particle speeds. (a) The original shape at a spin period of 2.26 hr. (b) The deformed shape.

The key finding from these efforts is that Didymos's failure mode strongly depends on its structural and material properties. While such a response is poorly constrained for Didymos, when subject to an external destabilizing effect, this possibly sensitive body may experience material flows on its surface and/or in the interior on various scales. For the DART impact, such an external effect may result from the falling of the DART impact–driven ejecta from Dimorphos, if the ejecta cloud carries enough kinetic energy to surface materials on Didymos.

Whether the falling ejecta could indeed cause Didymos's reshaping on any scales is another issue. Because of uncertainties about the system's geophysical properties and the DART impact conditions, it is difficult to predict the probabilities. Earlier work using a DEM model predicts that if Didymos's surface cohesive strength is greater than 5 Pa, collisions of ejecta particles with diameters of 10 cm and speeds of 1 m s$^{-1}$ do not change Didymos's surface conditions; therefore, no significant reshaping may be expected (Sánchez & Scheeres 2018). Since Didymos's escape speed due to gravity is only about 1.0 m s$^{-1}$, only a limited amount of fast ejecta particles would travel directly to Didymos and have an effect. However, the net effect of falling ejecta on surface disturbance is not well understood.

Furthermore, Didymos may have experienced an impact event with the DART impact energy every 0.4 Myr (Section 2.5). Impact events exceeding $Q_D^*$ for Didymos may occur every >10 Gyr. If this is the case, Didymos's surface and structure should possess proper conditions that can resist disturbance by the falling of DART-driven ejecta to avoid complete collapse. The key issue, however, is that granular behaviors and slow impact processes in microgravity are likely different from those observed on a terrestrial surface (Garcia et al. 2015; Murdoch et al. 2017; Brisset et al. 2019; Sánchez & Scheeres 2021). Therefore, such disturbing processes could act differently on Didymos, which needs further quantitative assessment.

Next, there is a question of how to measure the reshaping magnitudes. Didymos reshaping as driven by DART-driven falling ejecta can be inferred by using Earth-based telescopes to measure Didymos's spin change. With an anticipated timing precision of <0.1 s by the end of 2023 (Pravec & Scheirich 2018), measurement of shape changes as small as ~1 cm will be possible (Rivkin et al. 2021). Additional evidence of perturbation to Didymos could include the development of relatively large dust clouds after the DART impact, compared to the DART impact–driven ejecta alone, implying mass shedding from Didymos. If this happens, Didymos's rapid spin may be lessened, as ejected particles may take away some of the asteroid's angular momentum. Dimorphos reshaping may also be determined by measuring the total mass of ejecta from optical measurements by Earth-based telescopes and LICIACube. The determined ejecta mass will be converted to the crater cavity volume. Combining this volume with information about Dimorphos's pre-impact shape and the impact site location from proximity imaging can constrain its post-impact shape.

Finally, also note the possibility that accretion of ejecta onto Didymos and Dimorphos may change the bodies' spin states directly. Low-speed ejecta do not immediately escape from the system—previous work shows that the impact speeds of these trapped particles range from ~10 cm s$^{-1}$ at Dimorphos to ~60 cm s$^{-1}$ at Didymos (Yu & Michel 2018; Rossi et al. 2022). The momentum carried by the ejecta, while tiny compared to the rotational angular momenta of the main bodies, may have a non-negligible cumulative effect on the asteroids' spins. Preliminary work indicates that Didymos may experience a spin period change of up to a few seconds via ejecta accretion.

Recent efforts to analyze this effect (Rossi et al. 2022) indicate that the uncertainties on the total excavated mass from the impact crater—and, hence, the total reimpacting mass—make it challenging to estimate how secondary impacts will affect the asteroids' spin states. Should the effect be non-negligible, it would be necessary to find a way to distinguish the spin change due to ejecta accretion alone from that due to any subsequent reshaping process. Thus, it is necessary to quantify the minimum ejecta mass landing on either body that will induce a non-negligible change in the period. A detailed evaluation is underway.

### 5.2.3. Reshaping-induced Mutual Orbit Perturbation

When the DART spacecraft collides with Dimorphos, surface particles will be ejected from the impact site with various speeds, and some of them may reach Didymos (e.g., Richardson & O'Brien 2016; Yu et al. 2017; Yu & Michel 2018; Raducan & Jutzi 2022). While the amount of ejecta and the total energy input to this body are not well constrained, if such a physical perturbation on the surface is strong enough, a reshaping process may occur on Didymos, due to its short spin period. If this is the case, this process will change the gravity field, leading to a non-negligible perturbation in the mutual interactions between Didymos and Dimorphos (Hirabayashi et al. 2017, 2019).

If any reshaping of Didymos is large enough, the resulting changes to the mutual orbit may be measurable. Due to Didymos's fast rotation, the shape change will likely lead to a more oblate shape and a larger $J_2$ moment. Increasing it without changing Didymos's mass will shorten the orbital period. Thus, assessments without accounting for this effect could underestimate $\beta$. Figure 12 shows an example of Didymos's reshaping and the resulting orbital period change as a function of the magnitude of the reshaping along the $z$-axis, i.e., the spin axis. If collisions of the DART-induced ejecta with Didymos cause net deformation along the $z$-axis of 0.7 m ($D_0 - D$ in Figure 12(a)), which is consistent with earlier work (Hirabayashi et al. 2017, 2019), it would make the mutual orbit period 7.8 s shorter than the original one (Hirabayashi et al. 2022; Nakano et al. 2022), which exceeds the DART measurement requirement (7.3 s; Rivkin et al. 2021). If the shape change is 4 m, the mutual orbit period change increases up to 45 s. At this level of reshaping, the mutual orbit period change linearly correlates with the magnitude of Didymos's reshaping, just as it does for Dimorphos's reshaping (Section 5.1.2).

DART impact outcomes could help constrain Didymos's surface/internal structure. If DART-driven reshaping happens to Didymos and is not negligible, then it also induces a non-negligible mutual orbit perturbation. This would imply that Didymos has a relatively weak structure (either on the surface or in the interior) and frequently experiences reshaping processes driven by meteoroid impacts and other disturbances, controlling the magnitude of the reshaping. Such processes induce mass shedding, which could contribute to the formation and evolution of Dimorphos (e.g., Walsh et al. 2008). On the other hand, if the DART impact–driven ejecta do not cause reshaping of Didymos, a possible interpretation is that Didymos has a relatively strong mechanical structure and lacks weak material layers; otherwise, the surface should fail structurally because of the 2.26 hr spin period. In this case, any weak materials would have been depleted during the system's early formation stage, and thus the evolution process would be less dynamic at present.

## 6. Secular Evolutionary Effects

The dynamical state of binary systems evolves due to gravitational and nongravitational perturbations, such as tides and BYORP. BYORP is the binary-YORP effect, whereby a tidally locked satellite's thermal emission contributes to a net torque on the mutual orbit, leading to secular changes in the mutual orbit's semimajor axis and eccentricity (Ćuk & Burns 2005). The BYORP coefficient depends only on the secondary's shape, but the overall evolution of the system is tightly coupled with the attitude state of the secondary and the tidal strength (Ćuk & Burns 2004; Ćuk & Nesvorný 2010; McMahon & Scheeres 2010a, 2010b; McMahon 2014, 2016). The effects of tides on a binary system mainly cause a change in the semimajor axis through the transfer of angular momentum between the bodies. Further, energy dissipation inside both bodies leads to the circularization of the orbit and the damping of free modes. These secular effects of tides are characterized by the ratio of the quality factor $Q$, accounting for the energy dissipation, to the Love number $k_2$, associated with the additional potential due to tidal deformation (Goldreich & Soter 1966; Murray & Dermott 2000), which depends on the internal structure and rheology of the asteroid.

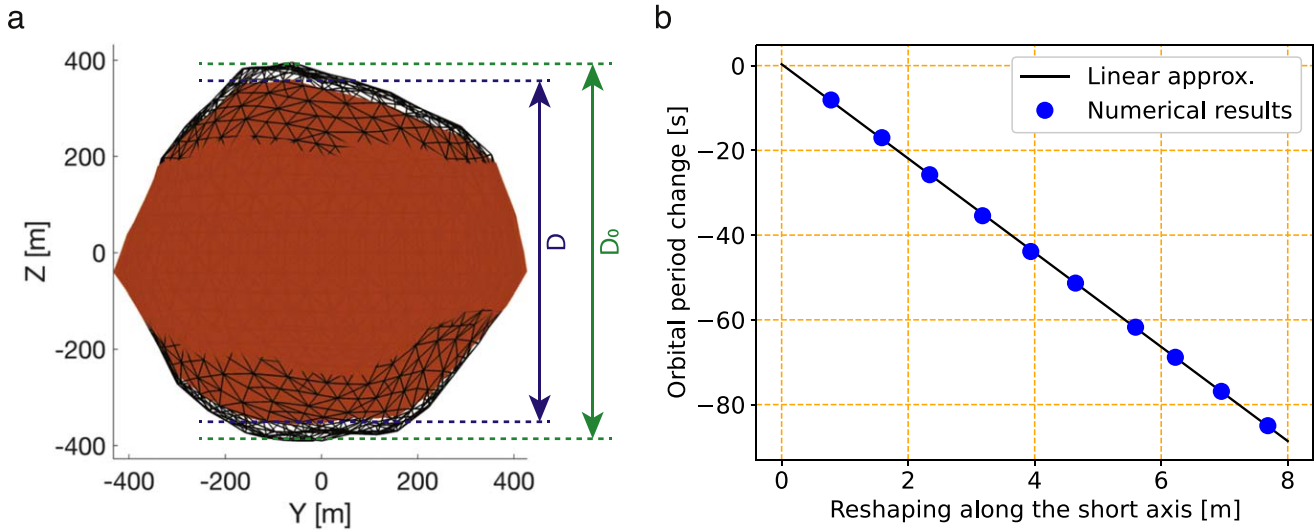

**Figure 12.** Didymos's deformed shape and the resulting orbit period change as a function of the reshaping magnitude. (a) An example of reshaping making the shape more oblate along the maximum principal axis (corresponding to the spin axis, here the *z*-axis). The *y*-axis corresponds to the intermediate axis. (b) Orbital period change as a function of deformation along the leading direction. The blue dots are derived from the FEM approach and the black line gives a linear approximation.

The parameters of BYORP and the tides affecting the Didymos system are currently poorly constrained. The shape of the secondary, necessary for determining the BYORP coefficient, is currently unknown (the DRA assumes an ellipsoidal shape—see Section 2.2). The internal structures and rheologies of the asteroids are not known either, and the resulting estimations of tidal parameters of other asteroids vary by orders of magnitude. McMahon et al. (2016) simulated the possible dynamical evolution of Didymos based on the modeled (not observed) BYORP coefficient for the 66391 Moshup–Squannit (formerly 1999 $KW_4$) secondary shape model (Ostro et al. 2006). When scaled to the Didymos–Dimorphos size and state, this resulted in a possible orbit semimajor axis expansion/contraction rate of $\pm 1.66$ cm yr$^{-1}$, where the sign can be changed by flipping the secondary orientation 180° around its spin axis. This semimajor axis rate translates to a quadratic mean anomaly drift rate of $\mp 2.8°$ yr$^{-2}$, or a period drift of 0.91 s yr$^{-1}$.

The estimation of the secular changes in the semimajor axis of the Didymos system due to the combined effect of BYORP and tides, using historical observations, yields a mean anomaly drift with $3\sigma$ uncertainty $\Delta M_d = 0.15 \pm 0.14$ deg yr$^{-2}$ (Agrusa et al. 2021; Scheirich & Pravec 2022). This rate is quite small, with the $3\sigma$ uncertainty near zero, meaning that any secular changes to the orbit are slow; however, it indicates that the BYORP effect must currently be acting, as the orbit is shrinking to cause the positive mean anomaly drift. Unfortunately, this does not provide information with which to constrain the current BYORP coefficient, since it only indicates that the BYORP effect is slightly stronger than tidal expansion. However, based on current tidal models (Jacobson & Scheeres 2011) with an expected $k_2/Q \approx 10^{-6}$, the BYORP effect at Dimorphos is fairly weak. This could imply that the current shape of Dimorphos is much more symmetric than Squannit (similar to the discussion in Scheirich et al. 2021) or that Dimorphos's attitude dynamics are otherwise limiting BYORP (Ćuk et al. 2021; Quillen et al. 2022). If the tides are stronger than suspected, the BYORP coefficient for Dimorphos could be similar to, or even larger than, that of Squannit. Alternatively, Dimorphos may be in or very close to a BYORP–tide equilibrium, a state predicted analytically by Jacobson & Scheeres (2011), in which the tides and BYORP effectively cancel each other (Agrusa et al. 2021).

The DART mission is expected to provide the shapes and hence the BYORP coefficient for the Didymos system, which will later be improved by the Hera mission. After the DART impact, how the secular evolution of the Didymos system may progress is highly uncertain. Until the arrival of Hera, secular changes to the semimajor axis are expected to be rather small, based on the abovementioned estimate, unless the BYORP effect is shut off by the barrel instability triggered following the DART impact (Section 3.2). However, if the attitude is simply excited, the BYORP effect can continue to act. Furthermore, the shape of Dimorphos will necessarily change, as discussed in Section 5, which will directly change the BYORP coefficient. The DART impact could also impart some small inclination change to the binary orbit, which would further complicate the BYORP predictions, since current models typically assume zero inclination. For $k_2/Q \approx 10^{-5}$, and without the BYORP effect acting, the change of the semimajor axis due to tides will be on the order of a few cm over 5 yr, which is challenging but not impossible to detect with ground-based observations. Hera will constrain the mass as well as the dynamics of the binary system, with the aim of the submeter determination of the semimajor axis (Hera Internal Mission Requirements Document ESA-HERA-TECSH-RS-022793, v. 2.10). In addition, Hera instruments will help to reveal the interior structure, composition, and surface thermal properties, as well as the past and future evolution of the Didymos system.

Table 5
Post–DART Impact Dynamical Effects with Predicted Ranges and Observables

| Excitation | Motion | Range ($\beta = 1$–$2$) | Observable |
|---|---|---|---|
| Mutual orbit | Orbit period change (see Figure 13(b)) | −8.8 min to −17 min (see Table 2) | Mean anomaly drift |
| Orbit period variation | Orbit period fluctuations (see Figure 13(b)) | Tens of seconds to several minutes (see Figures 3(a) and 4) | Nonconstant orbit period/mean anomaly drift |
| Dimorphos libration state, aligned | Yaw excitation (see Figure 13(d)) and eccentricity (see Figure 13(b)) | Depends on Dimorphos's shape (see Figure 6) | Dimorphos's lightcurve and, possibly, comparisons between DART and Hera imaging |
| Dimorphos libration state, nonaligned | Roll and pitch excitation (see Figures 13(e) and 13(f)) | Depends on Dimorphos's shape (see Figure 6) | Dimorphos's lightcurve and, possibly, comparisons between DART and Hera imaging |
| Dimorphos shape modification | Orbit period change (see Figure 13(b)) and Dimorphos spin excitation (see Figure 13(d)) | Depends on Dimorphos's shape (up to ∼200 s; see Figure 10) | Dimorphos's lightcurve and, possibly, comparisons between DART and Hera imaging |
| Didymos shape modification | Orbit period change (see Figure 13(b)) and Didymos spin change | Depends on Didymos's shape (see Figure 12) | Didymos's lightcurve and, possibly, comparisons between DART and Hera imaging |

## 7. Conclusions and Future Outlook

The DART spacecraft is on its way to impact Dimorphos—the satellite of the binary near-Earth asteroid system Didymos—at around 23:14 UTC on 2022 September 26. LICIACube will detach from DART several days before impact to image the crater, ejecta, and unseen hemisphere of the moon. Dimorphos's orbital period change will be measured by ground-based assets over subsequent weeks, providing an estimate of the momentum transfer enhancement factor, $\beta$. Together, this will be the first test of the kinetic impactor strategy for deflecting a potentially hazardous asteroid. As part of the AIDA cooperation between NASA and ESA, Hera will rendezvous with the system several years later to characterize it fully. The combination of knowledge from DART, LICIACube, and Hera will provide the first fully documented impact and deflection experiment at actual asteroid scale.

This paper collects the current understanding of the dynamics of the Didymos system and the best-informed expectations of how the dynamics will be affected by the DART impact.

1. Didymos is an S-class asteroid that is likely an escaped member of the Baptistina family (Section 2.1). Observations of the Didymos binary system are consistent with a low-eccentricity, dynamically relaxed state (Section 2), but high-order gravity potential terms are significant at the 1% level, leading to the expectation of low-amplitude orbital and attitudinal oscillations (Section 2.6). Observations to date also cannot rule out a recent impact or gravitational perturbation that could put the system into an excited state at encounter (Section 2.5), including Dimorphos's nonsynchronous rotation, but this is considered unlikely. An accurate prediction of the system's orbital phase at encounter is not possible from dynamics modeling alone, due to uncertain initial conditions, but observations do constrain the target location within mission requirements (Agrusa et al. 2020).

2. The value of $\beta$ resulting from the DART impact is expected to lie in the range from 1 to 5, based on impact modeling (Stickle et al. 2022), and it will be estimated using Equation (1) from Rivkin et al. (2021). $\beta$ is most sensitive to the measured change in Dimorphos's orbital period, but it will have an uncertainty dominated by the poorly constrained mass of the target. An independent measure of its heliocentric counterpart, $\beta_\odot$, may be achievable via occultation observations and could aid in constraining $\beta$ and/or the physical properties of Dimorphos (Section 3.4).

3. The Didymos system is sensitive to dynamical instabilities that depend on Dimorphos's poorly constrained shape (Section 3.2). These may be triggered prior to encounter (see the earlier point) or by the DART impact, leading to libration, with amplitude dependent on $\beta$. A rolling or "barrel" instability may also be present prior to impact or be triggered by the impact. Regardless of Dimorphos's shape, libration must be induced if Dimorphos is tidally locked, since the post-impact orbital period will be less than the moon's rotation period. In addition, the post-impact orbital period will fluctuate by anywhere on the order of tens of seconds to several minutes, on timescales ranging from a few days to months, depending on the target shape and $\beta$, as a result of a periodic exchange of angular momentum between Dimorphos's spin state and the mutual orbit (Section 3.3). Given a more accurate shape model of Dimorphos post-impact, if observations are able to characterize the orbit period variations, an additional constraint on $\beta$ and the libration amplitude may be obtainable.

4. The pre- and post-impact dynamics presented in this paper were modeled using the rigid-body integrator GUBAS (Davis & Scheeres 2020). The dynamical state most consistent with observations and a relaxed initial condition is generated any time there is a change in the Didymos DRA. Post-impact steady-state models are also generated for a range of assumed $\beta$ values (Sections 2.6 and 3.1). This modeling makes the well-supported assumption that the momentum transfer is short compared to the orbital period (Section 3.5).

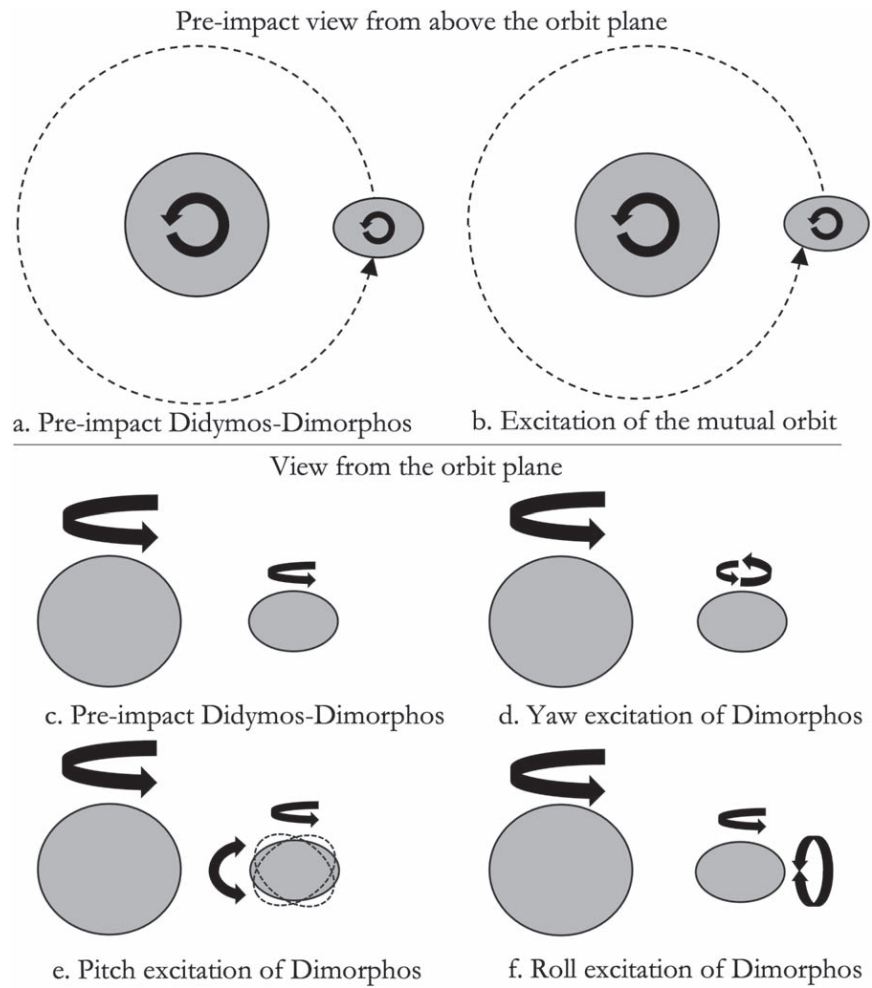

**Figure 13.** Summary cartoons of dynamical excitation due to the DART impact. Panels (a) and (c) show a representation of the relaxed dynamics of the Didymos–Dimorphos system from a view above the orbit plane (a) and a view in the orbit plane (c). The DART impact is expected to excite the mutual orbit directly (b), causing the orbital period to change both in an average sense (by shrinking the orbit, in this case) and with fluctuations about the average (eccentric motion). It will likely also excite the libration of Dimorphos about different rotation axes, including: (d) yaw excitation, oriented along the rotation axis of Dimorphos—fluctuations in the rate of rotation during a given rotation and/or a change in the average rotation rate; (e) pitch excitation, oriented in and out of the orbit plane—a nodding of the face of Dimorphos from the perspective of Didymos; and (f) roll excitation, oriented about the axis connecting Dimorphos and Didymos—a rocking of the face of Dimorphos from the perspective of Didymos.

5. Models in which either component of Didymos is treated as a rubble pile are broadly consistent with the rigid-body modeling (Section 4; H. Agrusa 2022, in preparation), but they do reveal the sensitivity of the primary to shape deformation due to its near- or supercritical rotation state (Section 5, Hirabayashi et al. 2022). A change of shape of either body due to the impact has implications for the final orbital period and, therefore, $\beta$. In the case of an ejecta-induced shape change to Didymos, the effect on $\beta$ can in principle be constrained by measuring a change in the primary rotation period, although how to distinguish this from spin-up due to accretion alone remains to be investigated. Dimorphos's shape change effect will be constrained by converting the measured total ejecta mass to the crater size. Estimations of the impact histories of both bodies suggest that the primary at least is robust against DART-scale impacts, however (Section 2.5).

6. Although observations constrain any secular effects on the system's orbital period to be small on the timescale of the DART mission, both the BYORP effect and mutual gravity tides are likely play a role in the long-term evolution of the system (Section 6). The relevant BYORP and tide parameters are poorly known at present, but the upper bounds on secular acceleration from observations to date are consistent with a fairly symmetric secondary shape, stronger tidal dissipation than expected from current models, and/or the possibility that the system is close to a BYORP–tide equilibrium. Post-impact observations from the ground and by Hera may help distinguish between these possibilities.

For reference, Table 5 summarizes the principal dynamical effects discussed that may be triggered by the DART impact. The motion of each excitation is shown schematically in Figure 13, for ease of understanding each of the particular motions described in the paper. The range of each excited motion is also provided either in the table directly, for a modest range of $\beta$, or via reference to the appropriate figure. Last, we list the principal observable associated with each excitation.

The true aftermath of the DART impact will be determined in the weeks, months, and then years following the event, leading to new research assessing the kinetic impactor deflection experiment. No doubt there will be surprises!

We thank the anonymous reviewers whose comments improved the manuscript. This work was supported in part by the DART mission, NASA Contract #80MSFC20D0004 to JHU/APL. F.F. acknowledges funding from the Swiss National Science Foundation (SNSF) Ambizione grant No. 193346. O.K., A.C.B., I.G., M.J., P.M., S.D.R., K.T., and Y.Z. acknowledge funding support from the European Union's Horizon 2020 research and innovation program under grant agreement No. 870377 (project NEO-MAPP). J.W.M. acknowledges support from the DART Participating Scientist Program (#80NSSC21K1048). S.R.S. acknowledges support from the DART Participating Scientist Program (#80NSSC22K0318). A.C.B. acknowledges funding from the Spanish MICINN RTI2018-099464-B-I00. E.D. and A.R. acknowledge financial support from Agenzia Spaziale Italiana (ASI, contract No. 2019-31-HH.0 CUP F84I190012600). E.G.F. acknowledges that some of this work was carried out at the Jet Propulsion Laboratory, California Institute of Technology, under a contract with NASA (#80NM0018D0004). P.M. acknowledges support from the French space agency CNES. R.N. acknowledges support from NASA/FINESST (NNH20ZDA001N).

## ORCID iDs

Derek C. Richardson ● https://orcid.org/0000-0002-0054-6850
Harrison F. Agrusa ● https://orcid.org/0000-0002-3544-298X
William F. Bottke ● https://orcid.org/0000-0002-1804-7814
Andrew F. Cheng ● https://orcid.org/0000-0001-5375-4250
Siegfried Eggl ● https://orcid.org/0000-0002-1398-6302
Fabio Ferrari ● https://orcid.org/0000-0001-7537-4996
Masatoshi Hirabayashi ● https://orcid.org/0000-0002-1821-5689
Jay McMahon ● https://orcid.org/0000-0002-1847-4795
Stephen R. Schwartz ● https://orcid.org/0000-0001-5475-9379
Ronald-Louis Ballouz ● https://orcid.org/0000-0002-1772-1934
Adriano Campo Bagatin ● https://orcid.org/0000-0001-9840-2216
Elisabetta Dotto ● https://orcid.org/0000-0002-9335-1656
Oscar Fuentes-Muñoz ● https://orcid.org/0000-0001-5875-1083
Seth A. Jacobson ● https://orcid.org/0000-0002-4952-9007
Martin Jutzi ● https://orcid.org/0000-0002-1800-2974
Rahil Makadia ● https://orcid.org/0000-0001-9265-2230
Alex J. Meyer ● https://orcid.org/0000-0001-8437-1076
Patrick Michel ● https://orcid.org/0000-0002-0884-1993
Ryota Nakano ● https://orcid.org/0000-0002-9840-2416
Guillaume Noiset ● https://orcid.org/0000-0002-1649-7176
Sabina D. Raducan ● https://orcid.org/0000-0002-7478-0148
Alessandro Rossi ● https://orcid.org/0000-0001-9311-2869
Paul Sánchez ● https://orcid.org/0000-0003-3610-5480
Daniel J. Scheeres ● https://orcid.org/0000-0003-0558-3842
Stefania Soldini ● https://orcid.org/0000-0003-3121-3845
Angela M. Stickle ● https://orcid.org/0000-0002-7602-9120
Paolo Tanga ● https://orcid.org/0000-0002-2718-997X
Kleomenis Tsiganis ● https://orcid.org/0000-0003-3334-6190
Yun Zhang ● https://orcid.org/0000-0003-4045-9046